%% file: main.tex
\documentclass[journal]{IEEEtran}
\usepackage{amsmath,amsfonts,amssymb}
\usepackage{mathtools}
\usepackage{array}

\usepackage[ansinew]{inputenc}
\usepackage{bm}
\usepackage{subcaption}	
\usepackage{import}
\usepackage[squaren]{SIunits}
\usepackage{upgreek}
\usepackage{hyperref}
\usepackage{physics}
\usepackage{cite}

\renewcommand{\figurename}{Fig.}
\renewcommand{\tablename}{Tab.}
\newcommand{\capname}{Sec.}

\newcommand{\figref}[1]{\figurename~\ref{#1}}
\newcommand{\tabref}[1]{\tablename~\ref{#1}}
\newcommand{\capref}[1]{\capname~\ref{#1}}

\usepackage{graphicx}

\usepackage{tikz}
\usepackage{algpseudocode}
\usepackage{algorithm}
\usepackage{multirow}
\usepackage{subcaption}

\begin{document}

\definecolor{mygreen}{RGB}{0,105,0}

\title{Steady-State Error Compensation in Reference Tracking and Disturbance Rejection Problems for Reinforcement Learning-Based Control}
\author{Daniel~Weber,
        Maximilian~Schenke
        ~and~Oliver~Wallscheid,~\IEEEmembership{Member,~IEEE}
\thanks{D. Weber and M. Schenke  are with the Department of Power Electronics and Electrical Drives while O. Wallscheid is with the Department of Automatic Control, Paderborn University, 33098 Paderborn, Germany e-mail: \{weber, schenke, wallscheid\}@lea.upb.de.}
}

\maketitle

\begin{abstract}
Reinforcement learning (RL) is a promising, upcoming topic in automatic control applications.
Where classical control approaches require a priori system knowledge, data-driven control approaches like RL allow a model-free controller design procedure, rendering them emergent techniques for systems with changing plant structures and varying parameters.
While it was already shown in various applications that the transient control behavior for complex systems can be sufficiently handled by RL, the challenge of non-vanishing steady-state control errors remains, which arises from the usage of control policy approximations and finite training times.
To overcome this issue, an integral action state augmentation (IASA) for actor-critic-based RL controllers is introduced that mimics an integrating feedback, which is inspired by the delta-input formulation within model predictive control.  
This augmentation does not require any expert knowledge, leaving the approach model free. 
As a result, the RL controller learns how to suppress steady-state control deviations much more effectively.
Two exemplary applications from the domain of electrical energy engineering validate the benefit of the developed method both for reference tracking and disturbance rejection. 
In comparison to a standard deep deterministic policy gradient (DDPG) setup, the suggested IASA extension allows to reduce the steady-state error by up to 52\,$\%$ within the considered validation scenarios.
\newline

\end{abstract}


\input{Introduction}

\input{RL}

\input{DDGP}

\input{Envs}

\input{Results}

\input{Conclusion}

\input{acknowledgment}

\bibliographystyle{IEEEtran}
\bibliography{references}
\end{document}

%% file: Introduction.tex
\section{Introduction}
\label{cap:intro}
%





Reference tracking and disturbance rejection are the two main application fields of control engineering. They play an essential role for many industry branches and are the basis of modern automation systems. In this context. consider a dynamic discrete-time system \eqref{eq:dynamic_system}. 
The goal of reference tracking and disturbance rejection is to control the system in such a way that the system's output $\bm{y}$ is equal to a desired reference signal $\vb*{y}^*$ even if disturbances are present:
\begin{alignat}{3}
&  && \underset{\bm{u}}{\text{min}} \: \Vert \bm{e}_k \Vert_p \nonumber\\
\mbox{s.t.} \hspace{0.4cm} & \bm{e}_k  && = \bm{y}_k^* - \bm{y}_k \nonumber\\
& \bm{x}_{k+1}   && = \bm{f}_k(\bm{x}_k, \bm{u}_k, \bm{d}_k) + \bm{\chi}_k, \label{eq:dynamic_system} \\
& \bm{x}_{k_0}&& = \bm{x}_\mathrm{0}, \nonumber \\
& \bm{y}_k &&  = \bm{h}_k(\bm{x}_k, \bm{u}_k) + \bm{\psi}_k. \nonumber 
\end{alignat}
Here, $\vb*{x} \in \mathcal{X} \subseteq \mathbb{R}^n$ represents the system state, $\vb*{u} \in \mathcal{U} \subseteq \mathbb{R}^m$ the system input, $\vb*{d} \in \mathcal{D} \subseteq \mathbb{R}^o$ the disturbance, $\vb*{y},\vb*{y}^* \in \mathcal{Y} \subseteq \mathbb{R}^r$ the output and its reference, $p \geq 1$ defines the used control error norm, $k$ denotes the time index, $\bm{x}_{\mathrm{0}}$ is the initial state of the system and $\bm{\psi}$ and $\bm{\chi}$ represent additive measurement and process noise, respectively\footnote{Bold symbols denote non-scalar quantities such as matrices and vectors as well as functions with a multidimensional value range.}.
The function
\mbox{$\vb*{f} : \mathcal{X} \times \mathcal{U} \times  \mathcal{D} \xrightarrow[]{} \mathcal{X}$} defines the state transition over time and 
\mbox{$\bm{g}: \mathcal{X} \times \mathcal{U} \xrightarrow[]{} \mathcal{Y}$} the output function defining the available measurement signals.

\subsection{State of the Art}
As stated in \cite{Levine2011}, both, the transient and the steady-state behavior are important characteristics of controller performance.
Standard feedback controllers, e.g., proportional-integral-derivative (PID) elements, are able to compensate the steady-state error for step-like reference tracking and disturbance rejection applications in linear time-invariant systems due to the stateful nature of the integral term \cite{Goodwin2000}. 
Additionally, a disturbance model can be taken into account during the controller design process to allow for feed-forward disturbance compensation.
Similarly, such a disturbance model can be considered in the domain of model predictive control (MPC) \cite{Rawlings2017}, \cite{Karamanakos2020} as presented in \cite{Borrelli2007}.
The disadvantage is that these methods typically require accurate system knowledge 
and necessitate a surplus of modelling effort each time the plant system changes, whereas data-driven control approaches like reinforcement learning (RL) can overcome this issue by employing a model-free approach that interacts with and learns from the plant system directly.
However, they introduce different challenges such as control accuracy and safety issues.
The latter is discussed for example in \cite{Mannucci2018}, \cite{Li2022}.
In the following this contribution will focus on the challenge of steady-state control accuracy during reference tracking and disturbance rejection applications.
Fields of applications can be, e.g.,
autopilots that follow a predefined course \cite{Kiran2021}, 
robot arms in industrial production lines\cite{Peters2006}, \cite{Inoue2017}, or
automated crane systems in harbours, train stations or warehouses \cite{Wang2017}.

Non-vanishing steady-state errors in RL applications arise from the usage of control policy approximations and finite training times.
Like shown in \cite{Kiumarsi2015} or \cite{Liu2015} a remaining steady-state error is often noticeable. 
As shown in \cite{Engel2014}, proper reward design allows to trade off between a fast transient response and a low steady-state error.
In the end, reward design alone can not sufficiently solve the problem of non-vanishing steady-state control error.  
Also, model-based RL approaches that deal with reference tracking problems have been suggested \cite{Alharkan2021}. A similar concept can be found in \cite{Kamalapurkar2017}, where only parts of the control plant have to be known. These approaches, therefore, come with the requirement of a model that needs to be available a priori or needs to be identified during operation. Moreover, steady-state error free control can be only achieved for special cases utilizing simplifying assumptions such as disturbance-free control plants. 

Several approaches in the literature combine classical control approaches and RL. 
For example, in \cite{Lu2021} an RL agent superimposes the output of a classical PID controller.
A similar approach is presented in \cite{Ye2021}, where a state feedback controller is extended with a feed-forward part adjusted by an RL agent to minimize the steady-state error. 
In \cite{Qin2018} the parameters of a classical PID controller are adjusted using an RL agent on a continuous action space. 
The previously mentioned articles have shown that the combination of both approaches can outperform the classical control approach as well as the standard RL agent when it is used on its own.
Instead of configuring conventional control approaches with data-driven methods, this paper will address the application of a pure RL controller that directly operates within the control loop and is augmented to tackle the issue of stationary control error without losing its capability of fast transients.

Steady-state accuracy is of exceptional importance in many application domains such as energy systems.
Therefore, the first exemplary scenario verifying the effectiveness of the suggested approach is a voltage control task for a grid-connected inverter with the task of a constant reference tracking facing disturbance rejection.
The second exemplary problem deals with a current control task of an electric drive with a variable reference value.
An RL-based attempt for the latter scenario has been presented in \cite{Schenke2020}, where the remaining challenge of stationary control error has not been addressed.
The transfer from simulation to real-world experiment was delivered in \cite{Book2021}, which underlines that RL control is making its path into practical application.
In the corresponding article, the RL agent lacks behind concerning the steady-state error compared to a classical PI controller but outperforms it in terms of total demand distortion. 
The measurement-based validation of the simulation setting used for the power grid application can be found in \cite{Weber2021}. 
As shown in \cite{Ernst2009}, RL is already applied in literature to solve power grid stability and power quality problems, but in a discrete state and action space without the usage of deep neural networks and without focusing on stationary accuracy of the reference tracking problem. 
In most investigations on power grid stability, steady-state conditions are assumed while the dynamics during transients are neglected \cite{Zhou2021}, \cite{Wang2020}.
In the power grid application of this contribution we will focus on the advantage in transient behavior applying an RL agent in a continuous state and action space while significantly reducing the steady-state error. 

\subsection{Contribution}
To improve the steady-state behavior in RL control applications, this contribution introduces an integral action state augmentation (IASA) for arbitrary actor-critic-based RL controllers, which is derived from the delta-input approach within model predictive control \cite{borrelli2017} and mimics an integrating feedback loop. 
The approach does not require any expert knowledge and IASA remains, therefore, model free.
The main contributions of this article can be summarized as follows:
\begin{itemize}
    \item IASA to suppress steady-state control deviations in actor-critic-based RL controllers,
    \item corresponding reward and feature engineering to enable expedient learning behavior in consideration of IASA,
    \item extensive validation results in a reference tracking example in a continuous action space of an electrical drive application and 
    \item extensive validation results in a disturbance rejection example in a continuous action space of a power grid application.
\end{itemize}

%% file: RL.tex
\section{RL Fundamentals}
\label{cap:RL_fundamentals}

In classical RL settings, an RL controller (referred to as agent) interacts repeatedly at each time step $k$ with the control plant (referred to as environment). 
\begin{figure}[H]
	\begin{center}
		\includegraphics[width=0.9\linewidth]{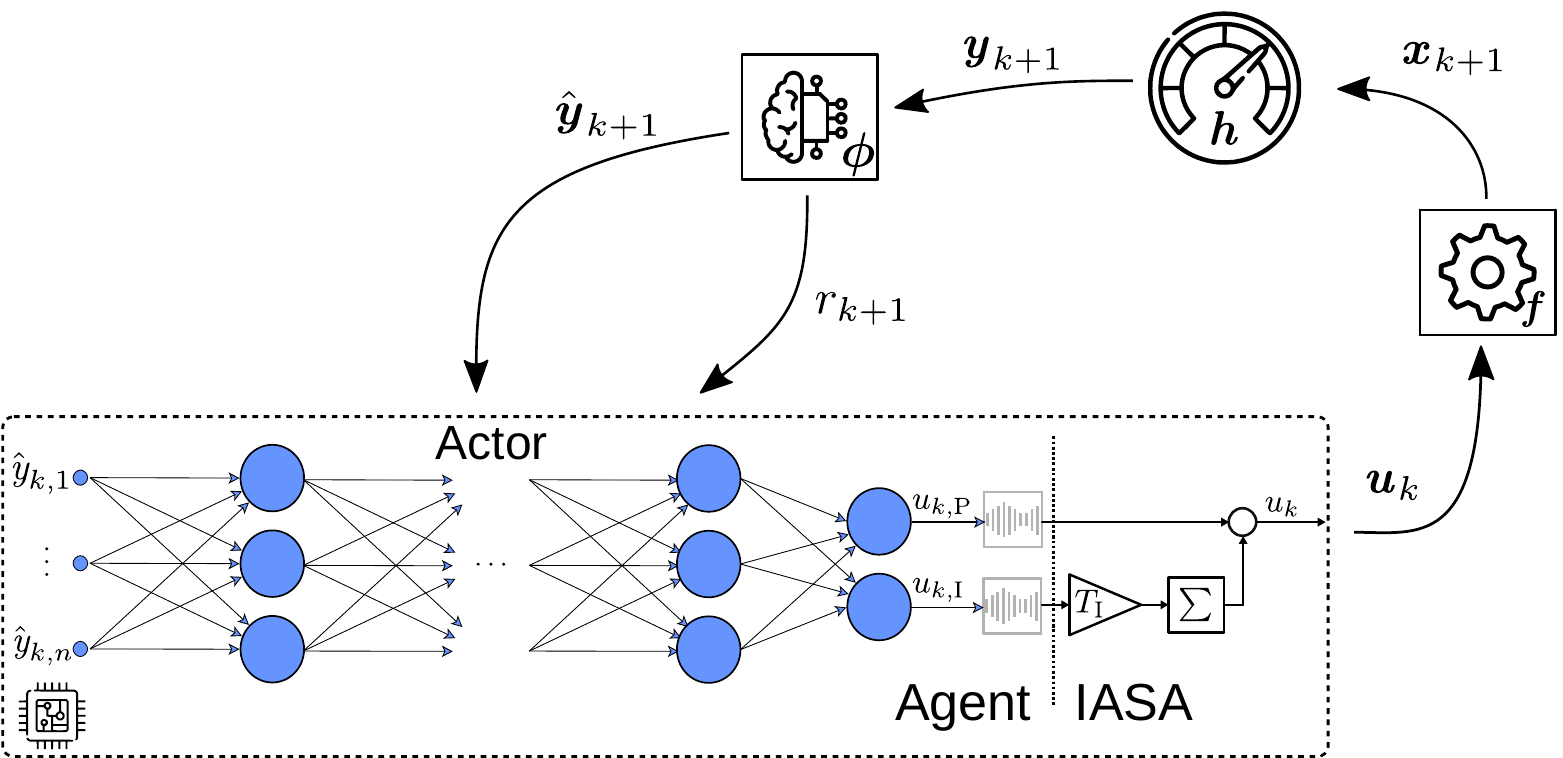}
		\caption{Steady-state error compensation (SEC) approach for an RL agent's actor interacting with an environment,  simplified for the case of a scalar actuating variable.}
		\label{fig:New_actor}
	\end{center}
\end{figure}
First we assume that all states are observable and are passed directly to the output.
Then, as can be seen in \figref{fig:New_actor}, the actor of a trained agent determines the control signal $\bm{u}_k$ based on the momentary environment state $\bm{x}_k$.
The applied action leads to the next state $\bm{x}_{k+1}$ based on the environment modeled as a Markov decision process (MDP) \cite{Sutton2018} which is defined by the tuple 
$\langle  \mathcal{X}, \mathcal{U}, \mathcal{P}, \mathcal{R}, \gamma\rangle$. 
Thereby, 
$\mathcal{X} \subseteq \mathbb{R}^n$ and $\mathcal{U} \subseteq \mathbb{R}^m$ define the possible state and action spaces, respectively. Moreover,
\begin{equation}
    \mathcal{P}(\bm{x}_{k+1}, \bm{x}_{k}, \bm{u}_k) = \mathrm{Pr}(\bm{x}_{k+1} | \bm{x}_{k}, \bm{u}_k)
\end{equation}
is the transition probability function of being in state $\bm{x}_{k}$ and taking the action $\bm{u}_k$ leading to the state $\bm{x}_{k+1}$, mapping \mbox{$\mathcal{P}: \mathcal{X} \times \mathcal{X} \times \mathcal{U} \xrightarrow[]{} [0,1]$}.
In the end, the action is rated by the reward $r_{k+1}$ given by a reward function \mbox{$\mathcal{R}:\mathcal{X} \times \mathcal{U} \xrightarrow[]{} \mathbb{R}$}.
From a control engineering perspective, the reward signal is representing the intended control objectives and a proper design can enhance the learning performance considerably.
$\gamma \in [0, 1[$ representing the discount factor which defines how far-sighted the agent is in pursuing a particular goal.

In an MDP the transition from state $\bm{x}_{k}$ to state $\bm{x}_{k+1}$ is only depending on the momentary state: 
\begin{equation}
    \mathrm{Pr}(\bm{x}_{k+1} | \bm{x}_{k}, \bm{u}_k) = \mathrm{Pr}(\bm{x}_{k+1} |  \bm{x}_{0},..., \bm{x}_{k}, \bm{u}_k)
\end{equation}
which is known as the Markov property.
If this does not hold, the environment is labeled partially observable, yielding a partially observable MDP (POMDP).
This can be the case, if, e.g., the described environment in \eqref{eq:dynamic_system} is subject to a significant, exogenous disturbance.
Then an explicit distinction must be made between state $\bm{x}_{k}$ and output $\bm{y}_{k}$ as described in detail in \capref{cap:DDPG}.

%

When applying RL to a control problem as shown in \figref{fig:New_actor}, the overall goal is to find the optimal policy that maps each state to an action which maximizes the return $g_k$, which is the expected cumulative reward:
\begin{equation}
\begin{split}
    g_k = &  \mathrm{E}_{}\{G_k | \bm{X}_k = \bm{x}_k, \bm{U}_k = \bm{u}_k\} \\
    = & \mathrm{E}_{}\{\sum_{i=0}^{\infty} \gamma^{i} R_{k+i+1} | \bm{x}_k, \bm{u}_k \}.
\end{split}
\label{eq:return}
\end{equation}
Here, $\mathrm{E}\{G_k\}$ denotes the expected value of the return, capital letters denote random variables, while lower case letters denote their realizations.
Using \eqref{eq:return}, the goal of the algorithm can be reformulated utilizing the Bellman optimality equation: 
\begin{equation}
    q^*(\bm{x}_k, \bm{u}_k) = \mathrm{E}\{R_{k+1} + \gamma \underset{\bm{u}_{k+1}}{\text{max}} \, q^*(\bm{x}_{k+1}, \bm{u}_{k+1}) | \bm{x}_k, \bm{u}_k\},
\end{equation}
wherein \mbox{$q(\bm{x}_k, \bm{u}_k)$} defines the action value function which maps the action $\bm{u}_k$ in state $\bm{x}_k$ to the expected return \mbox{$q: \mathcal{X} \times \mathcal{U} \xrightarrow[]{} \mathbb{R}$}.
To determine the best applicable action, a policy function $\bm{u}_k=\boldsymbol{\pi}(\bm{x}_k)$ with $\boldsymbol{\pi}: \mathcal{X} \xrightarrow[]{} \mathcal{U}$ needs to be learned that satisfies 
\begin{equation}
    \underset{u_{k}}{\text{max}}\, q(\bm{x}_{k}, \bm{u}_{k}) = q(\bm{x}_{k}, \boldsymbol{\pi}^*(\bm{x}_{k})).
\end{equation}

%% file: DDGP.tex
\section{Steady-State Error Compensation (SEC) in Reinforcement Learning}
\label{cap:MDP}
In the following, an integral action state augmentation (IASA) is presented to
 enable steady-state error compensation (SEC) for an actor-critic-based RL control algorithm. 
 
 \subsection{Steady-State Error Compensation}
\label{cap:SEC}
As discussed in \capref{cap:intro}, steady-state errors are an open task in RL control applications.
Approaches already exist in context with MPC control techniques, like shown in \cite{Borrelli2007}, including the disadvantage that a proper model knowledge is necessary.
In classical control, stateful systems like PID controllers are used to handle this issue.
Besides the proportional and derivative part the integrator term acts as memory, enabling the controller to reduce the steady-state error to zero for step-like disturbances or reference changes (for many system classes).
Contrary, most RL algorithms favor utilization of stateless feed-forward artificial neural networks (ANN). 
The obvious attempt -- the utilization of stateful ANN like recurrent neural networks or long short-term memory neurons -- is hardly compatible with the set of state-of-the-art learning algorithms. Instead, this article will present the augmentation of the agent's output with a stateful integral term, which can be implemented into any available actor-critic algorithm without requiring further changes to the learning mechanism.

\subsubsection{Integral Action State Augmentation}
As illustrated in \figref{fig:New_actor}, the integral term can be interpreted as an additional environment state from the agent's point of view.
The discrete-time integrator is implemented as a summation with weighting factor
$T_\mathrm{I} \in \mathbb{R}^+ $. 
This time constant is typically equal to the sampling time as described in \cite{Mattavelli2006}, but for the featured application it is set up as a hyperparameter.
Applying IASA configuration doubles the number of output neurons within the actor (one additional integral term per output).
As exemplary shown in \figref{fig:New_actor} for two output neurons, each output neuron pair will route one neuron to the environment and the other to the integrator unit.
Comparable to classical PI controllers, the output of the integrator is then added to the output of the first neuron, resulting in the applied action $\bm{u}_k$ :

\begin{equation}
    \bm{u}_{k}=
    \underbrace{
    \begin{bmatrix}
        u_{k,\text{P},1}\\
        \vdots\\
        u_{k,\text{P},m}
    \end{bmatrix}
    }_{\bm{u}_{k,\text{P}}}
        + 
        T_\text{I} \sum_{i=0}^{k} 
        \underbrace{
        \begin{bmatrix}
        u_{i,\text{I},1}\\
        \vdots \\ 
        u_{i,\text{I},m}
        \end{bmatrix}
        }_{\bm{u}_{k,\text{I}}},
        \label{eq:SEC-action}
\end{equation}
such that $\bm{u}$ may require clipping if the action space $\mathcal{U}$ is exceeded.
Whenever clipping needs to be applied, the internal state of the integrator unit $\bm{\zeta} = T_\text{I} \sum_{i=0}^{k} \bm{u}_{i,\text{I}}$ is reduced:

\begin{equation}
    \bm{\zeta} \leftarrow \bm{\zeta} + T_\mathrm{AW} \big( \text{clip}(\bm{u},-\bm{1},\bm{1}) - \bm{u} \big).
    \label{eq:sec_integrator}
\end{equation}
This compares to classical anti-windup measures (compare \cite{Goodwin2000}), with $T_\mathrm{AW} \in \mathbb{R}^+$ as a scaling parameter, typically chosen equal to $T_\text{I}$, but left as a free hyperparameter in the following.
Exploration noise, as it is used in many RL algorithms, needs to be superimposed to the action signals before the integrator unit as shown in \figref{fig:New_actor}.

\subsubsection{Reward Design}
According to a penalty function in MPC applications (cf. \cite{Rawlings2017}) a SEC-specific reward is used to enhance the learning process while interacting with the implemented integrator state: 
\begin{equation}
    r_{k,q} = \kappa_{k,q} \frac{1-\gamma}{m} \sum_{i=1}^{m} \left( - \sqrt{|\bm{u}_{k,q,i}|}\right),
    \label{eq:sec_reward}
\end{equation}
with $q \in \{\mathrm{P,I}\}$.
This reward needs to be incorporated into the task-specific reward design.
The factor $(1-\gamma)$ normalizes the discounted sum of rewards (see \capref{cap:DDPG}), limiting the action value to $[-1, 1]$ as described in \cite{Schenke2021}.  
$\kappa$ is a scaling parameter which is linearly reduced to zero starting from timestep $k_{\kappa_\text{P;I},0}$.
The additional reward acts as punishment for choosing an action, adjustable via the hyperparameters $\kappa$ and $k_{\kappa_\text{P;I},0}$.
Considering the usual PI controller from linear system control, the steady-state action is supposed to be handled solely by the integral term while the proportional term is only active during transients, which motivates the punishment of both terms individually. 

This SEC method can be applied to all state-of-the-art actor-critic methods, e.g., deep deterministic policy gradient (DDPG) \cite{Lillicrap2015}, twin delayed DDPG (TD3) \cite{Fujimoto2018}, soft actor-critic (SAC) \cite{Haarnoja2018}, trust region policy optimization (TRPO) \cite{Schulman2017} or 
proximal policy optimization algorithms (PPO) \cite{Schulman2017a}.
In the following, the DDPG algorithm will be used for the validation of the suggested approach, as it is a well-known candidate for problems that are defined on a continuous state and action space.

\subsection{Deep Deterministic Policy Gradient Algorithm with SEC Extension}
\label{cap:DDPG}
In order to take into account that not all MDP states may be observed directly or could be subject to measurement noise, a distinction is made between state $\bm{x}_k$ and output $\bm{y}_k$  in the following (compare \eqref{eq:dynamic_system}).
From a control engineering point of view, the system output $\bm{y}_k$ only covers the plant measurement. In comparison, the reference signal $\bm{y}^*_k$ needs to be added to the observed signals by proper feature engineering such that the RL  agent can actually learn its control task.
Using the feature function $\bm{\phi}: \mathcal{Y} \xrightarrow[]{} \hat{\mathcal{Y}} \subseteq \mathbb{R}^{\hat{r}}$ the resulting augmented signal $\hat{\bm{y}}_{k}$ is provided to the agent at each sampling instant $k$. Besides the reference information, further signals or signal processing can be part of $\bm{\phi}$.

As described in \capref{cap:RL_fundamentals} the goal is to find an optimal policy. 
To determine the best applicable action on a continuous state and action space 
the DDPG employs a deterministic policy function $\boldsymbol{\mu}$ with \mbox{$\boldsymbol{\mu}: \mathcal{Y} \xrightarrow[]{} \mathcal{U}$} that is learned in an off-policy fashion to maximize $q$, similar to \cite{Silver2014}. 
\begin{figure}[h!]
	\begin{center}
		\includegraphics[width=0.9\linewidth]{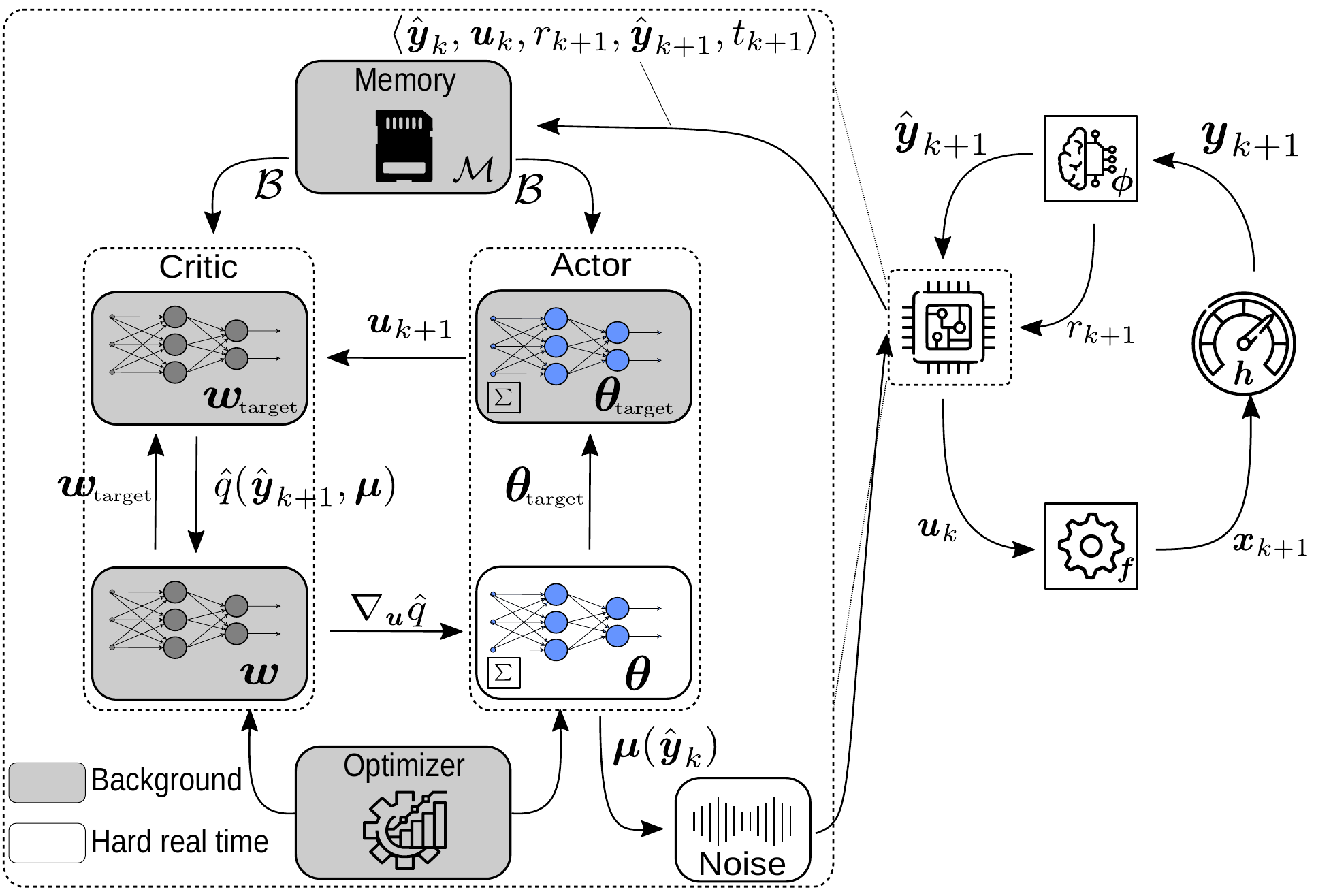}
		\caption{DDPG agent interaction with an environment. For more details about adapted SEC-actor see \figref{fig:New_actor}.}
		\label{fig:DDPG_plus_Integrator}
	\end{center}
\end{figure}
Therefore, the $q$ function needs to be differentiable with respect to the action, leading to a fitness function $J_{\boldsymbol{\mu}}$ that needs to be maximized:
\begin{equation}
\underset{\vb*{\theta}}{\text{max}} \, J_{\boldsymbol{\mu}}, \quad 
\mbox{s.t.} \, \, J_{\boldsymbol{\mu}} = \frac{1}{\mathcal{|B|}} \sum_{\hat{\bm{y}} \in \mathcal{B}} \hat{q}_{\vb*{\omega}}(\hat{\bm{y}}, \hat{\boldsymbol{\mu}}_{\vb*{\theta}}(\hat{\bm{y}})).\label{eq:J_mu}
\end{equation}
%
As shown in \figref{fig:DDPG_plus_Integrator}, ANN are used in most applications as function approximators for the policy, labeled $\hat{\boldsymbol{\mu}}_{\boldsymbol{\theta}}$ (with \mbox{$\hat{\boldsymbol{\mu}}_{\boldsymbol{\theta}}: \hat{\mathcal{Y}} \xrightarrow[]{} \mathcal{U}$}) and for the $q$ function, labeled $\hat{q}_{\vb*{\omega}}$ in the following (with \mbox{$\hat{q}_{\vb*{\omega}}: \hat{\mathcal{Y}} \times \mathcal{U} \xrightarrow[]{} \mathbb{R}$}).
Here, $\vb*{\theta}$ and $\vb*{\omega}$ represent the ANNs' adjustable parameters.
$\mathcal{B}$ is a batch of experience \mbox{$\textbf{\textit{b}} = \langle  \hat{\bm{y}}_k , \bm{u}_k , r_{ k + 1},  \hat{\bm{y}}_{k + 1}, t_{k+1} \rangle$}, while $t_{k+1} = 1$ if the state $\bm{x}_{k+1}$ is terminal, otherwise $t_{k+1} = 0$.
 
Next we need to train the approximator $\hat{q}_{\boldsymbol{\omega}}$.
The adjustable parameters $\vb*{\omega}$ are trained using the experience by minimizing the mean squared Bellman error:

\begin{align}
    &\underset{\vb*{\omega}}{\text{min}} \, J_q,  \quad  \,\mbox{s.t.} \, \, J_q= \frac{1}{\mathcal{|B|}} \sum_{\textbf{\textit{b}} \in \mathcal{B}} \bigg( \hat{q}_{\vb*{\omega}}(\hat{\bm{y}}_k, \bm{u}_k) - \label{eq:J_q}\\ 
    & \Big(r_{k+1} + \gamma (1-t_{k+1}) \, \hat{q}_{\vb*{\omega_\text{target}}}\big(\hat{\bm{y}}_{k+1}, \hat{\boldsymbol{\mu}}_{\vb*{\theta}_\text{target}}(\hat{\bm{y}}_{k+1})\big) \Big) \bigg)^2. 
\nonumber
\end{align}


Similar to deep $q$-networks (DQN) \cite{Mnih2015} some improvements stabilizing the learning process are used:

\subsubsection{Target networks}
the estimation target 
\begin{equation*}
    \Big(r_{k+1} + \gamma (1-t_{k+1}) \hat{q}_{\vb*{\omega_\text{target}}}\big(\hat{\bm{y}}_{k+1}, \hat{\boldsymbol{\mu}}_{\vb*{\theta}_\text{target}}(\hat{\bm{y}}_{k+1})\big) \Big)
\end{equation*} within the cost function \eqref{eq:J_q} that is needed for learning $\hat{q}$ is not calculated depending on the momentary trained parameters but on so-called target parameters ($\vb*{\theta}_\text{target}$ and $\vb*{\omega}_\text{target})$ \cite{Sutton2018}.
These target parameters are updated in a low-pass filter fashion on basis of the trained parameters
\begin{align}
    \begin{split}
        \vb*{\theta}_\text{target} &\xleftarrow[]{} (1- \tau) \vb*{\theta}_\text{target} + \tau \vb*{\theta}, \\
        \vb*{\omega}_\text{target} &\xleftarrow[]{} (1- \tau) \vb*{\omega}_\text{target} + \tau \vb*{\omega},
        \label{eq:target_update}
    \end{split}
\end{align}
wherein $\tau \in [0, 1]$ represents a damping parameter.
 
\subsubsection{Experience replay buffer} 
after each step the experience $\textbf{\textit{b}}$ is stored in the memory $\mathcal{M}$, which acts as a ring buffer of limited capacity, discarding the oldest data sample whenever a new one is added.
This allows the agent to draw uniformly distributed mini batches $\mathcal{B}$ of state transitions for an enhanced learning process because the temporal dependency in a mini batch of non-consecutive state transitions is greatly reduced.

\begin{algorithm}
\caption{DDPG pseudocode}
\label{alg:pseudocode}
\begin{algorithmic}
\State \textbf{Input:} diff. det. policy $\hat{\bm{\mu}}_{\bm{\theta}}(\hat{\bm{y}})$ and action-value fct.  $\hat{q}_{\bm{\omega}}(\hat{\bm{y}},\bm{u})$
\State Initialize \mbox{$\bm{\theta} = \bm{\theta}_{\text{target}}$}, \mbox{$\bm{\omega}_{\text{}} = \bm{\omega}_{\text{target}}$}, arbitrarily
\Repeat
\State observe $\bm{y}_k$ and add features  $\hat{\bm{y}}_k=\bm{\phi}(\bm{y}_k)$ 
\State apply $\bm{u}_k$ drawn from \eqref{eq:noise_action}
\State observe $r_{k+1}$, $\hat{\bm{y}}_{k+1}$ and $t_{k+1}$ 
\State store experience to replay buffer
\State \qquad $\mathcal{M} \leftarrow \langle  \hat{\bm{y}}_k , \bm{u}_k , r_{k+1},  \hat{{\bm{y}}}_{k+1}, t_{k+1} \rangle$
\State if ${\bm{x}}_{k+1}$ is terminal, reset environment
\If{time to update}
\State sample mini-batch $\mathcal{B} \subset \mathcal{M}$
\State update $\bm{\theta}$ by maximizing  $J_{\boldsymbol{\mu}}$ on $\mathcal{B}$, \eqref{eq:J_mu}
\State update $\bm{\omega}$ by minimizing $J_q$ on $\mathcal{B}$, \eqref{eq:J_q}
\State update weights of target networks based on \eqref{eq:target_update} 
\EndIf
\State $k \leftarrow k+1$
\Until{convergence condition is met}
\end{algorithmic}
\end{algorithm}

\subsubsection{Exploration noise}
as the trained policy is deterministic, action noise is added during the training process to enable the exploration of the state and action space. 
In the following a discrete-time Ornstein-Uhlenbeck (OU) noise 
\begin{equation}
    \boldsymbol{\nu}_k =  \boldsymbol{\nu}_{k-1} + \lambda (\eta - \boldsymbol{\nu}_{k-1}) \frac{\Delta_t}{\unit{1}{s}} + \sigma \sqrt{\frac{\Delta_t}{\unit{1}{s}}}\, D_\mathrm{N}(0,1)
    \label{eq:OUprocess}
\end{equation}
with zero mean is used, which is superimposed to the action determined by the policy 
\begin{equation}
\bm{u}_{k} = \hat{\boldsymbol{\mu}}_{\boldsymbol{\theta}}(\bm{y}_{k}) + \boldsymbol{\nu}_k.
\label{eq:noise_action}
\end{equation}
Herein, $\Delta_t$ defines the sampling time (normalized for sake of physical unit consistency), while $\lambda$, $\sigma$ and $\eta$ are the stiffness, diffusion and mean, respectively. $D_\mathrm{N}(0,1)$ denotes a realization from standard Gaussian random distribution.
Any necessary action clipping to accommodate for the limits of $\mathcal{U}$ is to be performed after this noise is added.

%% file: Envs.tex
\section{Application Examples}
This section presents the two exemplary applications in which the suggested SEC method is to be verified. Moreover, the DDPG parameterization  is featured.

\subsection{Electrical Power Grid Application}
\label{cap:OMG_setting}
\begin{figure}[htb]
	\begin{center}
		\includegraphics[width=0.9\linewidth]{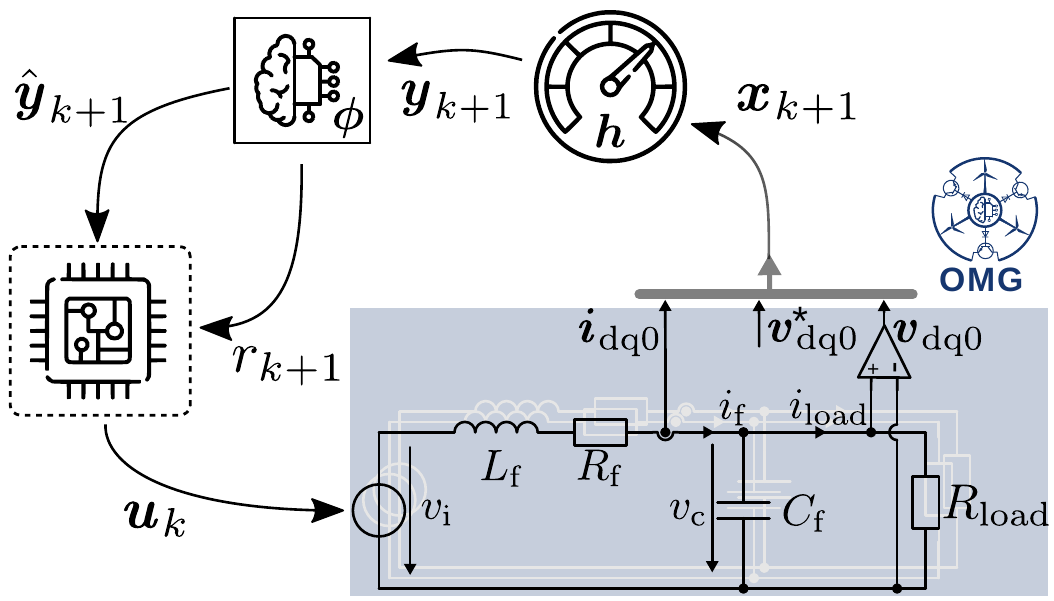}
		\caption{Power grid application setup under investigation.}
		\label{fig:OMG_setting}
	\end{center}
\end{figure}

As shown in \figref{fig:OMG_setting}, the first application under investigation is an electrical power grid fed by a voltage source inverter with a constant voltage reference and with significant disturbance that is based on the power intake of a resistive load. 
The scenario is characterized as a disturbance rejection problem with constant set-point reference and therefore modelled liked described in \capref{cap:MDP} as an POMDP.
This dynamically changing and unknown load $R_\text{load}$ is to be supplied by a sinusoidal voltage of \mbox{$v_\text{nom} = 120 \cdot \sqrt{2} \, \volt$} at $f = 60 \, \hertz$ in a common fixed three-phase reference system. 
The utilized power source is a three-phase inverter (fed by, e.g., a battery or a photovoltaic plant), which is connected to the load via an intermediate $LC$ line filter.
The changing load represents the unpredictable consumer behavior within the grid whose power demand unloads the employed line filter, therefore acting as a disturbance signal. An extensive description of the complete application and control plant setup as well as  executable simulation and training code can be found in \cite{Weber2021a}.

The control algorithm is implemented using the rotating dq0 reference frame, wherein the balanced sinusoidal grid voltages and currents can be represented as constants. This simplifies the problem definition as also the reference signals become constants.
This plant is simulated using the open-source OpenModelica Microgrid Gym (OMG) library \cite{OMG-code2020}, \cite{Bode2021} for Python \cite{python3_8}.
The transformation of the measured quantities in the fixed abc coordinate system to the rotating dq0 reference frame and inverse transformation of actions back to fixed abc frame is not shown in \figref{fig:OMG_setting} for the sake of simplicity. It is assumed to be a part of the environment such that the control agent never needs to handle any signals in the abc frame. 
Utilization of this transformation is a well-established method, as no further system information is necessary for applying this simplification \cite{Weber2021}, \cite{Mattavelli2006}.
A dead time of one simulation step is assumed within the plant system to account for the controller calculation delay that is common for digital control systems.
\begin{figure}
	\begin{center}
		\includegraphics[width=1\linewidth]{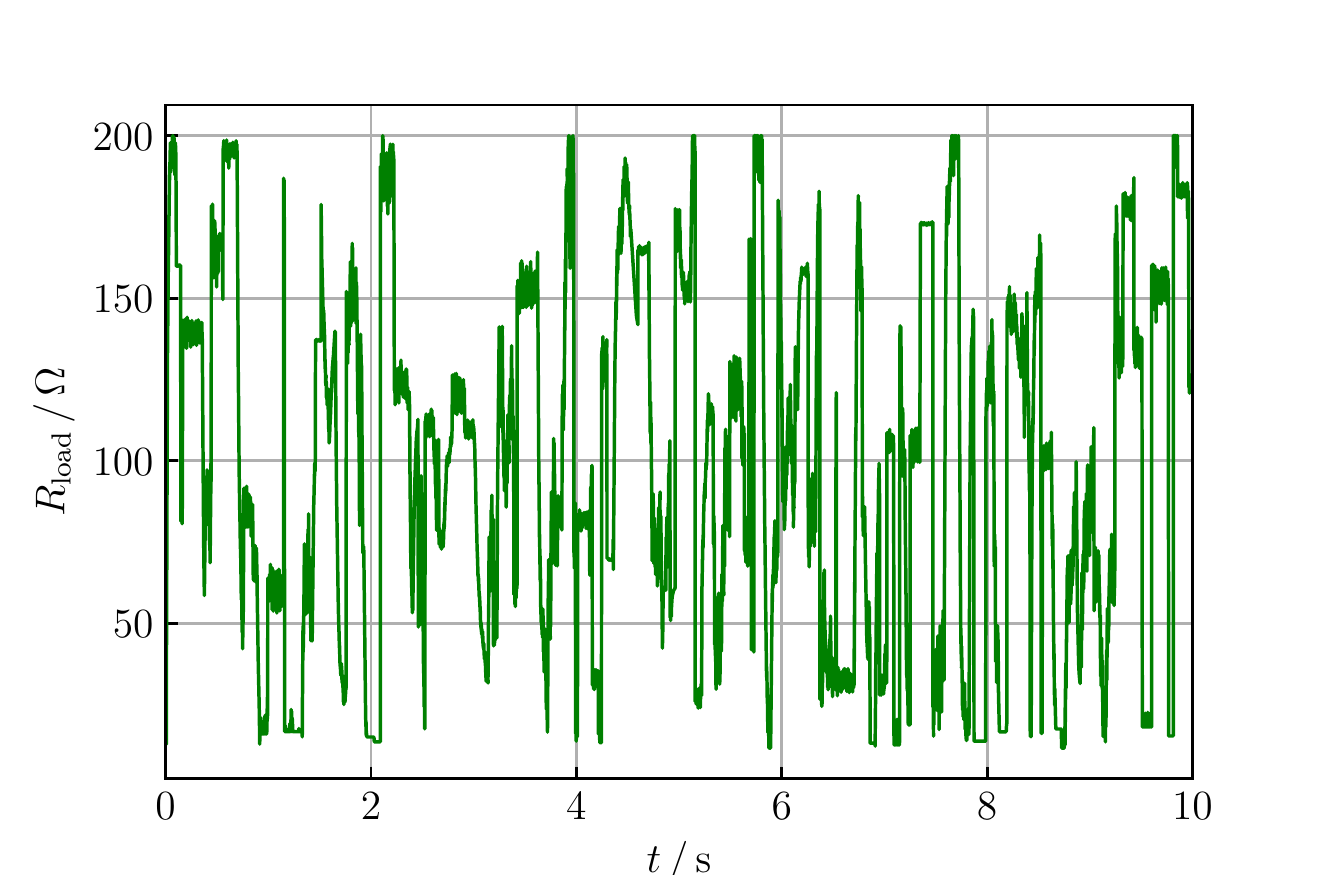}
		\caption{Exemplary power grid load curve used for validation.}
		\label{fig:OMG_R_load}
	\end{center}
\end{figure}
The same simulation configuration as described in \cite{Weber2021} is used in the following.
The output $\hat{\bm{y}}_k$ consists of the measurement signals depicted in \figref{fig:OMG_setting} ($\bm{v}_\mathrm{dq0}$, $\textbf{\textit{i}}_\mathrm{dq0}$) and a constant reference signal $\bm{v}_\mathrm{dq0}^*$ representing nominal grid operation regarding voltage amplitude and frequency.

The output is augmented within an interpreter to incorporate feature engineering that accelerates the learning process. This adds the following features:
\begin{itemize}
    \item the error between reference and measured voltages $\textbf{\textit{e}}_{v\mathrm{,dq0}} = \frac{1}{2} \left( \bm{v}_\mathrm{dq0}^* - \bm{v}_\mathrm{dq0} \right)$, 
    \item the past actions $\bm{u}_{k-1,\text{P}}$ and $\bm{u}_{k-1,\text{I}}$ that have been applied in the previous time step to account for the digital control delay, and
    \item the $\xi$ past voltage measurements $\bm{v}_{\mathrm{dq0}, \{k-\xi,...,k-1\}}$.
\end{itemize}
The latter provides additional history information to the control agent such that the POMDP structure of the problem (unknown grid load) can be implicitly handled. Accordingly, the featured output $\hat{\bm{y}}_k$ provided to the agent is defined by
\begin{equation}
    \begin{split}
       \hat{\bm{y}}_k = \bigg[ \frac{\textbf{\textit{i}}_{\mathrm{dq0, }k}^\mathsf{T}}{i_\mathrm{lim}} \, \, 
       \frac{\bm{v}_{\mathrm{dq0, }k}^\mathsf{T}}{v_\mathrm{lim}} \, \, 
       \frac{(\bm{v}_{\mathrm{dq0, }k}^*)^\mathsf{T}}{v_\mathrm{lim}} \, \, 
       \frac{\bm{e}_{v\mathrm{,dq0, }k}^\mathsf{T}}{v_\mathrm{lim}} \, \, 
       \frac{\bm{u}_{k-1,\text{P}}^\mathsf{T}}{\frac{v_\mathrm{DC}}{2}}\\ 
       \frac{\bm{u}_{k-1,\text{I}}^\mathsf{T}}{\frac{v_\mathrm{DC}}{2}} \, \,
       \frac{\bm{v}_{\mathrm{dq0, }\{k-\xi,...,k-1\}}^\mathsf{T}}{v_\mathrm{lim}} \bigg]^\mathsf{T}
       \label{eq:omg_x}
   \end{split}
\end{equation}
with $v_\mathrm{DC}$ representing the inverter's DC-link voltage. 
Furthermore, as usual for machine learning applications, each feature has been normalized to a range of $[-1,1]$ via division by the corresponding limit value.
Likewise, the action signals are interpreted as modulation indices for the inverter, which means that the agent's output is supposed to satisfy $\bm{u}_{\text{}k}\in[-1,1]^3$.

The reward for this application $r_{v,k+1}$ is computed using the mean root error (MRE) between the normalized measured voltages and the normalized reference values: 
\begin{equation}
    r_{k+1,v} =  - \frac{1-\gamma}{3} \sum_{{p \in \{\mathrm{d,q,0}\}}} \sqrt{\frac{\left|v_{k+1,p}^* - v_{k+1,p}\right|} {v_\mathrm{lim}}}.
    \label{eq:omg_reward_v}
\end{equation}
To account for both, the application goal and the specified SEC actor output, the task-specific reward and the SEC-specific \eqref{eq:sec_reward} have to be combined, leading to
\begin{equation}
    r_{k+1} = 
    \frac{r_{k+1,v} + r_\mathrm{k+1,P} + r_\mathrm{k+1,I}}{1+\kappa_\mathrm{k+1,P}+\kappa_\mathrm{k+1,I}}.
    \label{eq:reward_cases}
\end{equation}
The load is randomly altered within each time step, which is modeled using the OU-process described in \eqref{eq:OUprocess} with parameters $\lambda_k \in [10 , 1200]$, $\sigma_k \in [1 \, \Omega, 150 \, \Omega]$ and \mbox{$\eta_k \in [\text{\b{$R$}} = 14 \, \Omega, \bar{R} = 200 \, \Omega]$}, which are sampled randomly from a uniform distribution.
Moreover, with a probability of $0.2 \,\%$ in each step $k$ a load step is applied where new values for $\lambda_k$, $\sigma_k$ and $\eta_k$ are drawn.
This load step is with $50 \,\%$ probability either a step or a drift.
The new mean value is thereby represented by a sample $D_\mathrm{U}$ from a uniform distribution:
\begin{equation*}
    \eta_k = \text{clip}\big(D_\mathrm{U}(-10 \, \Omega , 200 \, \Omega),  \text{\b{$R$}} + D_\mathrm{N}(0\, \Omega,4 \, \Omega^2) , \bar{R} \big).
\end{equation*}
This configuration increases the occurrence of load values near the lower limit to enhance the learning during high power demand. 
Before they are applied the drawn load values are clipped to the upper mentioned range. 
To validate different agents, a $10 \, \second$ test case depicted in \figref{fig:OMG_R_load} was generated using the described stochastic process.

\subsection{Electric Motor Application}
\label{cap:GEM_setting}
\begin{figure}[hbt]
	\begin{center}
		\includegraphics[width=0.9\linewidth]{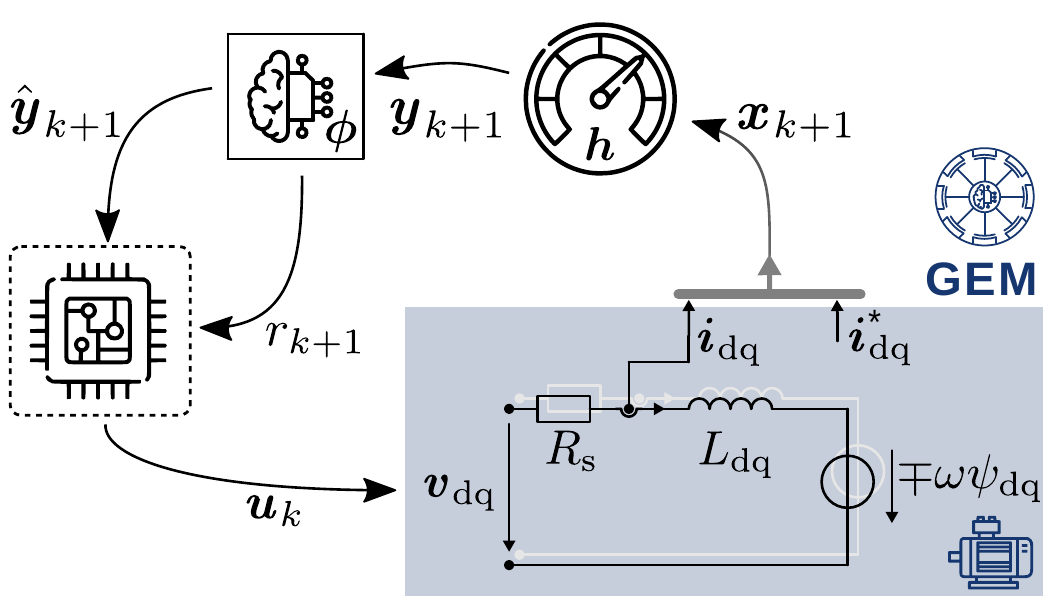}
		\caption{Electric motor application under investigation.}
		\label{fig:GEM_setting}
	\end{center}
\end{figure}

The SEC setup is further verified within an electric drive setting featuring a three-phase permanent magnet synchronous motor (PMSM) as depicted in \figref{fig:GEM_setting}.
Herein, the scenario is characterized as a reference tracking problem with a time-varying reference signal modelled as an MDP (cf. \capref{cap:MDP}).
Like in the previous example the complete application setting including executable simulation and training code can be also found in \cite{Weber2021a}.
The controlled variables are the motor currents in the above described rotating dq reference frame. 
Current control in electrical drives is a typical intermediate control problem that is to be solved to enable utilization within cascaded controllers for torque or speed control tasks.
The corresponding drive environment is simulated using the open-source  gym-electric-motor (GEM) \cite{Balakrishna2021} library.
During the training process, the reference values are drawn randomly at each sampling instant using the built-in reference generator. 
The output $\bm{y}_k$ consists of the current measurements and their corresponding reference values.
Similar to the power grid application the output is augmented via
\begin{itemize}
    \item the error between reference and measured currents $\textbf{\textit{e}}_{i\mathrm{,dq}} = \frac{1}{2} \left( \bm{i}_\mathrm{dq}^* - \bm{i}_\mathrm{dq} \right)$, 
    \item the actions $\bm{u}_{k-1,\text{P}}$ and $\bm{u}_{k-1,\text{I}}$ applied in the previous time step to account for the digital control delay, and
    \item the $\xi$ past current measurements $\bm{i}_{\mathrm{dq,}\{k-\xi,...,k-1\}}$.
\end{itemize}
The resulting featured output vector is defined by
\begin{equation}
    \begin{split}
       \hat{\bm{y}}_k = \bigg[ \frac{\textbf{\textit{i}}_{\mathrm{dq, }k}^\mathsf{T}}{i_\mathrm{lim}}  \, \, 
       \frac{(\bm{i}_{\mathrm{dq, }k}^*)^\mathsf{T}}{i_\mathrm{lim}} \, \, 
       \frac{\bm{e}_{i\mathrm{,dq, }k}^\mathsf{T}}{i_\mathrm{lim}} \, \, 
       \frac{\bm{u}_{k-1,\text{P}}^\mathsf{T}}{\frac{v_\mathrm{DC}}{2}}\\ 
       \frac{\bm{u}_{k-1,\text{I}}^\mathsf{T}}{\frac{v_\mathrm{DC}}{2}} \, \,
       \frac{\bm{i}_{\mathrm{dq, }\{k-\xi,...,k-1\}}^\mathsf{T}}{i_\mathrm{lim}} \bigg]^\mathsf{T},
       \label{eq:gem_x}
   \end{split}
\end{equation}
wherein $v_\mathrm{DC}$, again, defines the DC-link voltage of the inverter leading to an action space $\mathcal{U} = [-1, 1]^2$, with $\text{dim}(\mathcal{U}) = 2$, which means that also in this scenario the action signal can be understood as the modulation indices in the dq frame.

The task-specific reward is, again, defined utilizing the MRE between the reference and measured currents:
\begin{equation}
    r_{k+1,i} =  - \frac{1-\gamma}{2} \sum_{{p \in \{\mathrm{d,q}\}}} \sqrt{\frac{\left|i_{k+1,p}^* - i_{k+1,p}\right|} {i_\mathrm{lim}}}
    \label{eq:gem_reward_i}.
\end{equation}
The reformulation of this reward to account for the SEC setup is performed in the same way as presented in \eqref{eq:reward_cases}.

A test case of $10\,000$ steps is analyzed for validation.
Therein, new reference values are sampled per every $500$ steps. To create a representative reference value distribution, the DESSCA algorithm presented in \cite{Schenke2021a} is utilized. 

\begin{table*}[]
  \caption{Hyperparameter optimization setting and resulting best configuration set.}
  \label{tab:DDPG_hps}
  \renewcommand*{\arraystretch}{1.2}
  \centering
  \begin{tabular}[t]{l c r}
    \hline
    \textbf{Parameter}&\textbf{Search space}&\textbf{Best} \\
    \hline 
    training steps $M$ & &$5 \cdot 10^6$ (fixed)\\
    sampling time $\Delta_t$ & & $100 \cdot 10^{-6}\, \second$ \\
    episode steps $K$ & $\{1, ...,5000\}$ & $2811$\\
    activation function parameter $\beta_{\bm{\mu}}$ & $[1 \cdot 10^{-3}, 5 \cdot 10^{-1}]$ & $2.08 \cdot 10^{-1}$ \\
    activation function parameter $\beta_{q}$ & $[1 \cdot 10^{-3}, 5 \cdot 10^{-1}]$ & $6.79 \cdot 10^{-3}$\\ 
    hidden layers actor $l_{\bm{\mu}}$ & $\{1, ..., 4\}$ & 2 \\
    neurons per hidden layer $n_{\bm{\mu}}$ & $\{10, ..., 200\}$ & 25 \\
    hidden layers critic $l_{q}$ & $\{1, ..., 4\}$  & 4 \\
    neurons per hidden layer $n_{q}$ & $\{10, ..., 300\}$ & 295 \\
    learning rate $\alpha_\text{}$ & $[1\cdot 10^{-6}, 5\cdot 10^{-2}]$ & $3.75 \cdot 10^{-4}$ \\
    learning rate reduction start $k_{\alpha,0}$ & $\{0, ..., M\}$ & $1.375 \cdot 10^6$ \\
    learning rate reduction end $k_{\alpha,1}$ & $\{k_{\alpha,0}, ..., M\}$ & $1.62 \cdot 10^6$ \\
    final learning rate $\alpha_\text{final}$ & $[1\cdot 10^{-12}, \alpha]$ & $3.13 \cdot 10^{-4}$ \\
    \multirow{2}{*}{optimizer} & \{ADAM, SGD, & \multirow{2}{*}{ADAM}\\
    &RMSprop\} &\\
    \hline
  \end{tabular}
  \begin{tabular}[t]{l c r}
  \hline
    \textbf{Parameter}&\textbf{Search space}&\textbf{Best set}\\
    \hline
    discount factor $\gamma_\text{}$ &$[0.5, 0.999]$ & $0.946$ \\
    replay buffer size $\mathcal{M}$ & $\{2 \cdot 10^4, ..., M\}$ & $3.87 \cdot 10^6$ \\
    mini batch size $\mathcal{B}$ & $\{16, ..., 1024\}$ & $261$ \\
    target update parameter $\tau$ & $[1 \cdot 10^{-4}, 3 \cdot 10^{-1}]$ & $2.61 \cdot 10^{-3}$ \\
    weight scaling factor $\chi_{\bm{\theta},\text{w}}$ & $[5 \cdot 10^{-5}, 2 \cdot 10^{-1}]$ & $8.5 \cdot 10^{-4}$ \\
    bias scaling factor $\chi_{\bm{\theta},\text{b}}$ & $[5 \cdot 10^{-4}, 2 \cdot 10^{-1}]$ & $2 \cdot 10^{-2}$ \\
    penalty reduction start $k_{\kappa_\text{P},0}$ & $\{0, ..., M\}$ & $1.15 \cdot 10^6$ \\
    penalty reduction start $k_{\kappa_\text{I},0}$ & $\{0, ..., M\}$ & $2.75 \cdot 10^6$ \\
    noise stiffness $\lambda$ & $[1, 50]$ & $31.58$ \\
    noise diffusion $\sigma$ & $[1 \cdot 10^{-2}, 1]$ & $2.6 \cdot 10^{-2}$ \\
    training frequency $f_\text{train}$ & $\{1, 15 \cdot 10^3\}$ & $2$ \\
    reward scaling  $\kappa_\text{P}$ & $[0, 2]$ & $1.48$ \\
    reward scaling  $\kappa_\text{I}$ & $[0, 2]$ & $1.13$ \\
    integrator scaling  $T_\text{I}$ & $[5 \cdot 10^{-3}, 2]$ & 0.31 \\
    anti-windup scaling  $T_\text{AW}$ & $[1 \cdot 10^{-5}, 1]$ & 0.66 \\
    past measurements $\xi$ & $[0, 50]$ & 5 \\\hline
  \end{tabular}
\end{table*}

\subsection{DDPG Parameterization }
\label{cap:DDPH_HPs}
The verification of the SEC actor is conducted using the DDPG implementation of the Stable Baselines3 Python library \cite{stable-baselines3}.
The used hyperparameter (HP) setting for the training is specified in \tabref{tab:DDPG_hps}.
These parameters have been determined with use of the hyperparameter optimization framework Optuna \cite{Akiba2019} for the power grid example described in \capref{cap:OMG_setting}.
To validate the HP, the average reward of \eqref{eq:omg_reward_v} (with $\gamma = 0$) is calculated for $100000$ steps of the random load process described in \capref{cap:OMG_setting}. 
More then $12000$ samples of different HP were evaluated. 
The LeakyReLU activation $y = \text{max}(\beta x,x)$ is selected as activation function within every hidden layer.
While the actor uses the $\text{tanh(\cdot)}$ activation function in the output layer, inherently limiting the output range to $[-1,1]$, the critic features a linear output layer activation.
Both, actor and critic networks are trained using the same learning rate $\alpha$.

A linear learning rate scheduler is applied, reducing $\alpha$ to $\alpha_\text{final}$ over the course of step $k_{\alpha,0}$ to step $k_{\alpha,1}$.
For an enhanced training process the randomly chosen initial actor parameters are scaled using $\chi_{\bm{\theta},w}$ and $\chi_{\bm{\theta},b}$, respectively for weight and bias.
A parameter update is performed every $f_\text{train}$ steps.

%% file: Results.tex
\section{Results}
This section discusses the training and validation results that have been achieved for the previously introduced application scenarios.
The best set of hyperparameters listed in \tabref{tab:DDPG_hps} is utilized to train an SEC-DDPG agent using \cite{stable-baselines3}.
To rate the performance improvement, a standard DDPG (with identical hyperparameter setup) and a usual PI controller are investigated for a comparison.
The latter is able to eliminate steady-state error for step-like disturbance and reference signals by principle \cite{Goodwin2000}. Hence, it is a suitable baseline for steady-state control error investigations. 
To consider the influence of the random weight initialization, $550$ independent control agents are trained for each application.

\begin{figure}[htb]
\centering
 \includegraphics[width=1\linewidth]{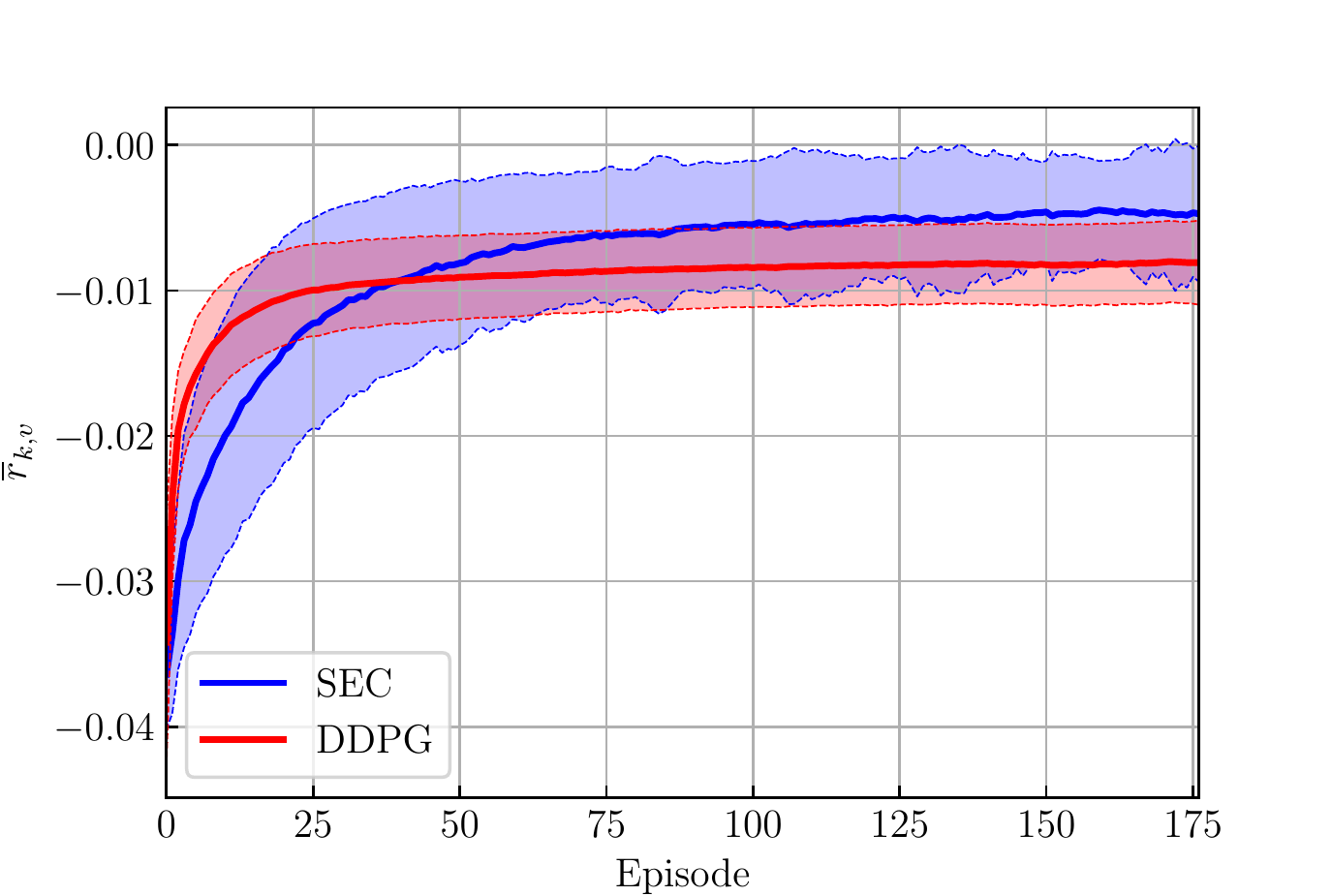}
\caption{Averaged mean reward per episode and the area of one standard deviation of the $550$ trained agents per approach for the electric power grid scenario.}
\label{fig:OMG_learning_curve}
\end{figure}

\subsection{Training Robustness}
In the following, the training behavior of the RL agents is investigated.
In order to evaluate the training in its entirety, no early stopping of the environments was applied during the training process. 
%
\subsubsection{Power Grid Application}
\figref{fig:OMG_learning_curve} shows the learning curve of the SEC and the bare DDPG approach for the power grid application during the training process.
The averaged mean reward per episode of the $550$ trained agents are shown.
Additionally a semi-transparent area of one standard deviation in both directions around the mean value is plotted.
The reward used during the training is calculated based on \eqref{eq:reward_cases} and the disturbance is calculated according to the random process based load described in \capref{cap:OMG_setting}. 
It can be deduced that the DDPG agent learns faster but is outperformed by the SEC agent at the end of the training process.
Furthermore, it can be observed that the SEC approach features a higher variance during the training process.
\begin{figure}[htb]
\centering
\includegraphics[width=1\linewidth]{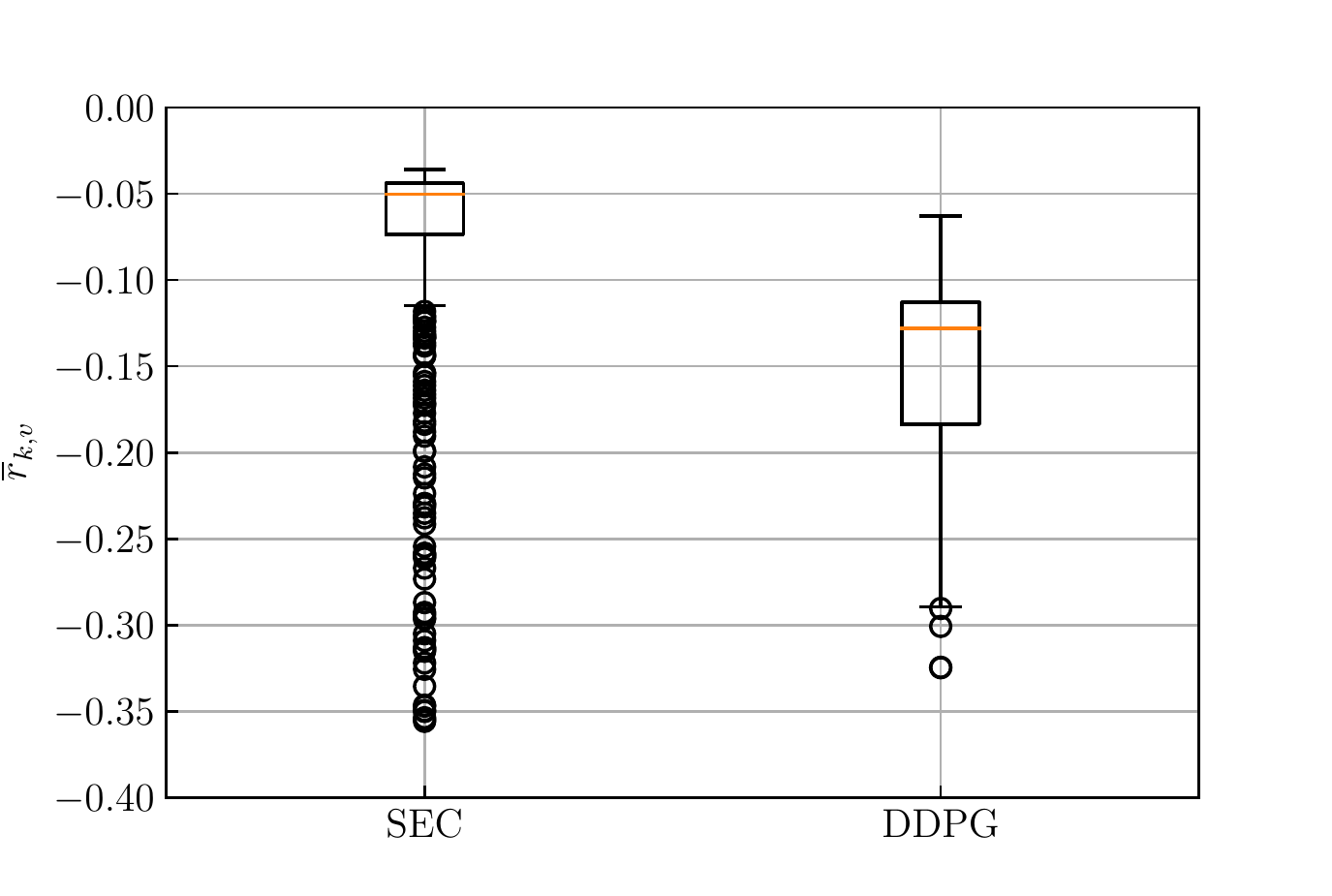}
\caption{Box plots of the control performance distribution for the power grid scenario based on a 100000 randomly generated steps test cases for 550 independently trained RL control agents each.}
\label{fig:OMG_boxplots}
\end{figure}
\figref{fig:OMG_boxplots} features box plots of the performance comparing the results of the test case of the standard DDPG and the SEC-DDPG agents for the power grid application over the investigated set of $550$ trained agents. 
Please note that the plot is cropped for an improved visibility of the relevant part, several very-low performing outliers of the noisier SEC approach are not depicted.
Within this test the reward normalization based on the discount factor is not used.
As visible within the plot, it can be concluded that the introduced SEC method significantly improves the average test case result.
Nonetheless, the amount of visible outliers as well as the variance within the learning curve indicate that the complexity of the training task increases, which complicates the controller training. This was to be expected in consideration of the additional state introduced with SEC's the integrating actor output.

\subsubsection{Electrical Drive Application}
\label{cap:drive_training}
\begin{figure}[htb]
\centering
\includegraphics[width=1\linewidth]{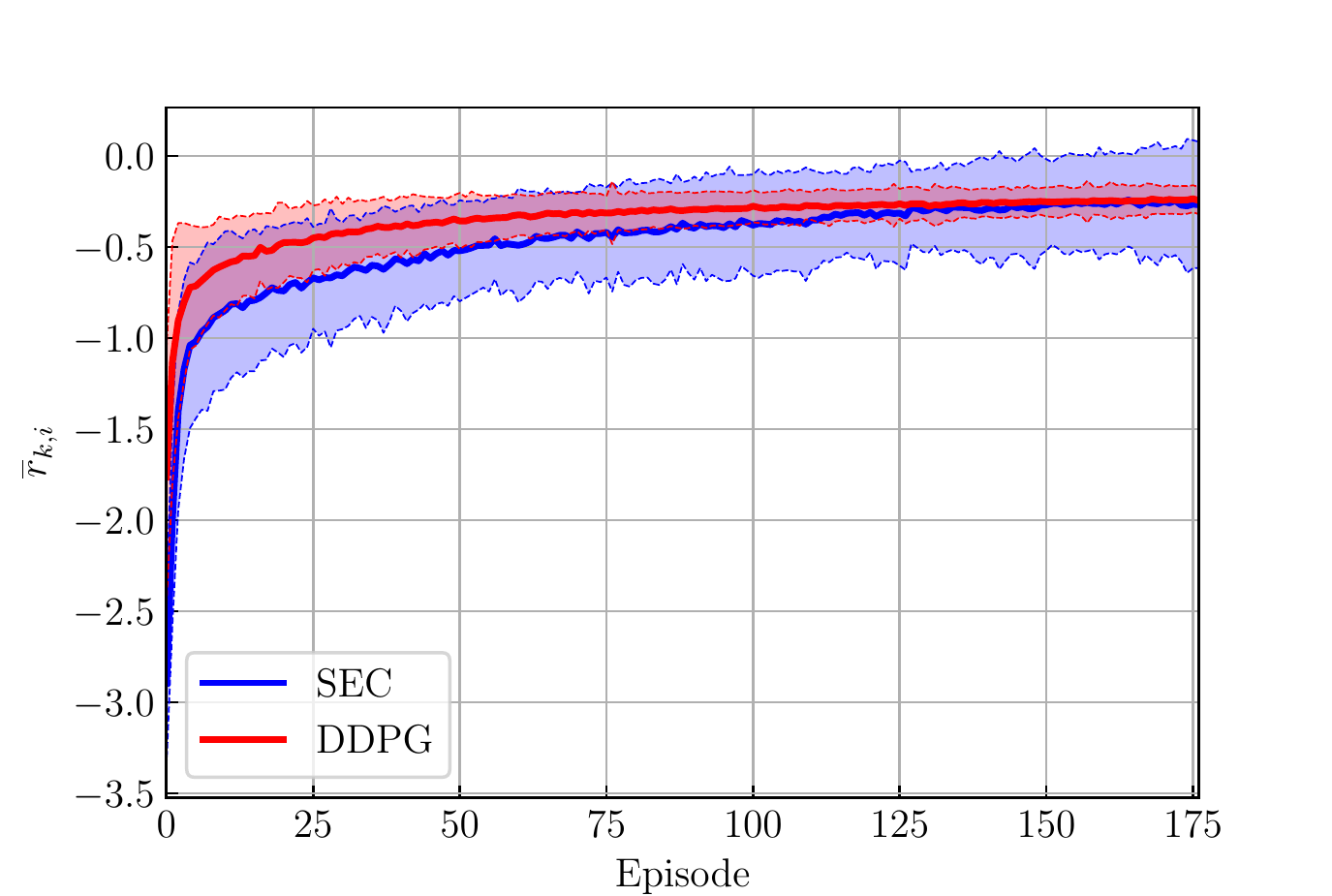}
\caption{Averaged mean reward per episode and the area of one standard deviation of the $550$ trained agents per approach for the electric drive scenario.}
\label{fig:GEM_learning_curve}
\end{figure}
A similar observation can be made with regard to the motor application reference tracking control task. 
\begin{figure}[htb]
\centering
\includegraphics[width=1\linewidth]{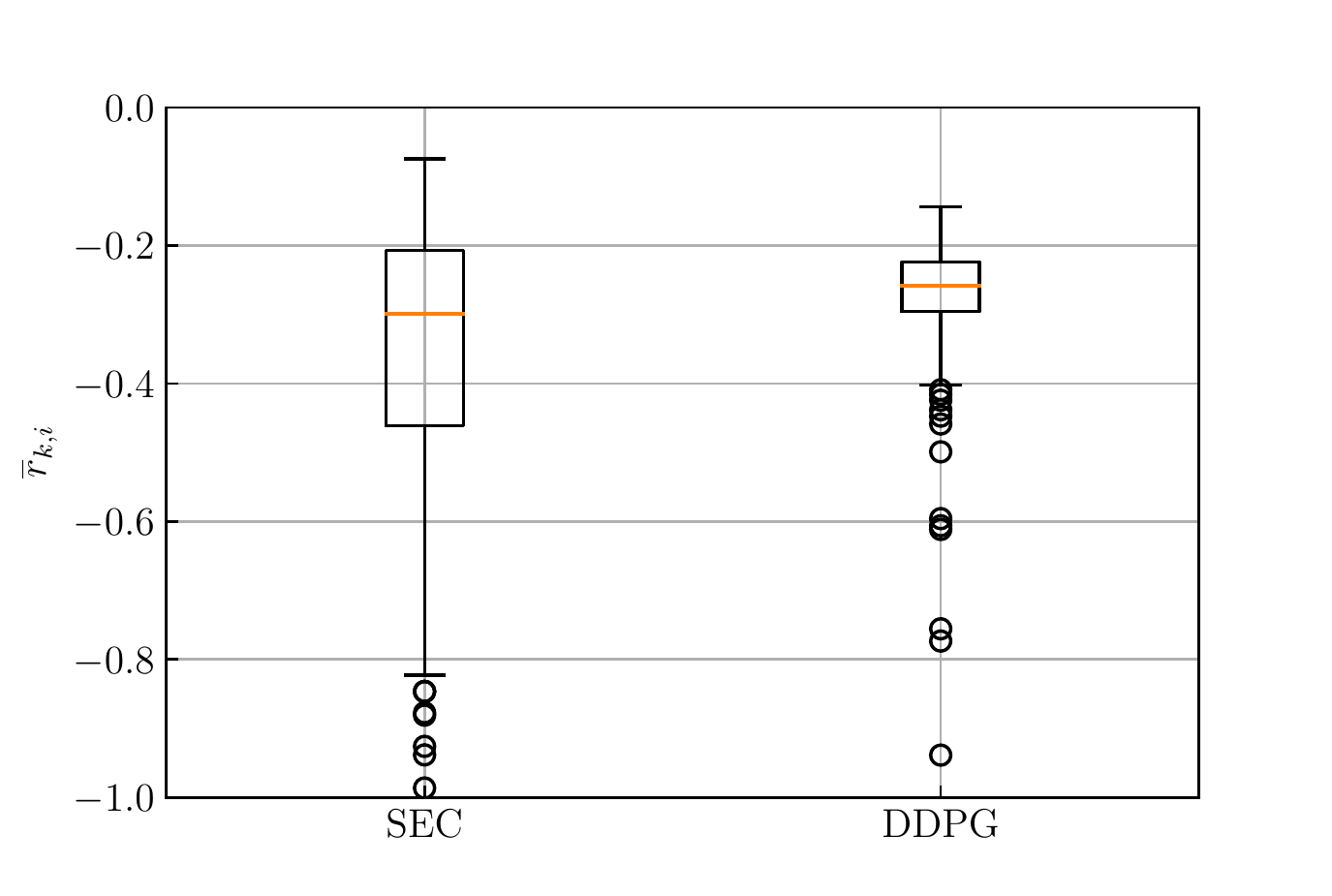}
\caption{Box plots of the control performance distribution for the electric drive scenario based the  $1 \, \mathrm{s}$ reference tracking test case for 550 independently trained RL control agents each.}
\label{fig:GEM_boxplots}
\end{figure}
\figref{fig:GEM_learning_curve} shows the corresponding learning curve during the training process.
Again, the averaged mean reward per episode and the area of one standard deviation of the $550$ trained agents per approach is shown.
The reward is calculated as described in \capref{cap:GEM_setting}.
It can be observed that the DDPG agent learns faster as already described in the power grid example.
At the end of the training process both approaches perform comparable, but the SEC approach outperforms the DDPG in the maximum.
Similar observations can be made in the corresponding box plots for the $550$ trained DDPG and SEC-DDPG agents as presented in \figref{fig:GEM_boxplots}.
The trained agents are applied to a $1 \,\mathrm{s}$ test case. 
Within the test case the reference values change every $50 \, \mathrm{ms}$.
Again, the diagram is cropped at the lower end for better visibility of the relevant area.
Concerning the best performing agents, the introduced method improves the control result visibly.
However, a large spread of the results, especially for the SEC application, is observable which in turn may be caused by the increasing complexity.
Additionally, it is noticeable that the median of the DDPG agent is a little better compared to the SEC-DDPG agent, but lacks behind in peak performance -- as already noticed in the training results.
Note that the hyperparameter tuning was based on the power grid application and, therefore, no optimized hyperparameter configuration was applied in this case.
An additional, application specific optimization can be expected to improve these results, especially in consideration of the newly introduced degrees of freedom that correspond to the SEC extension.

\begin{figure*}[htb]
\centering
\begin{subfigure}{.3\textwidth}
\centering
\includegraphics[width=1\linewidth]{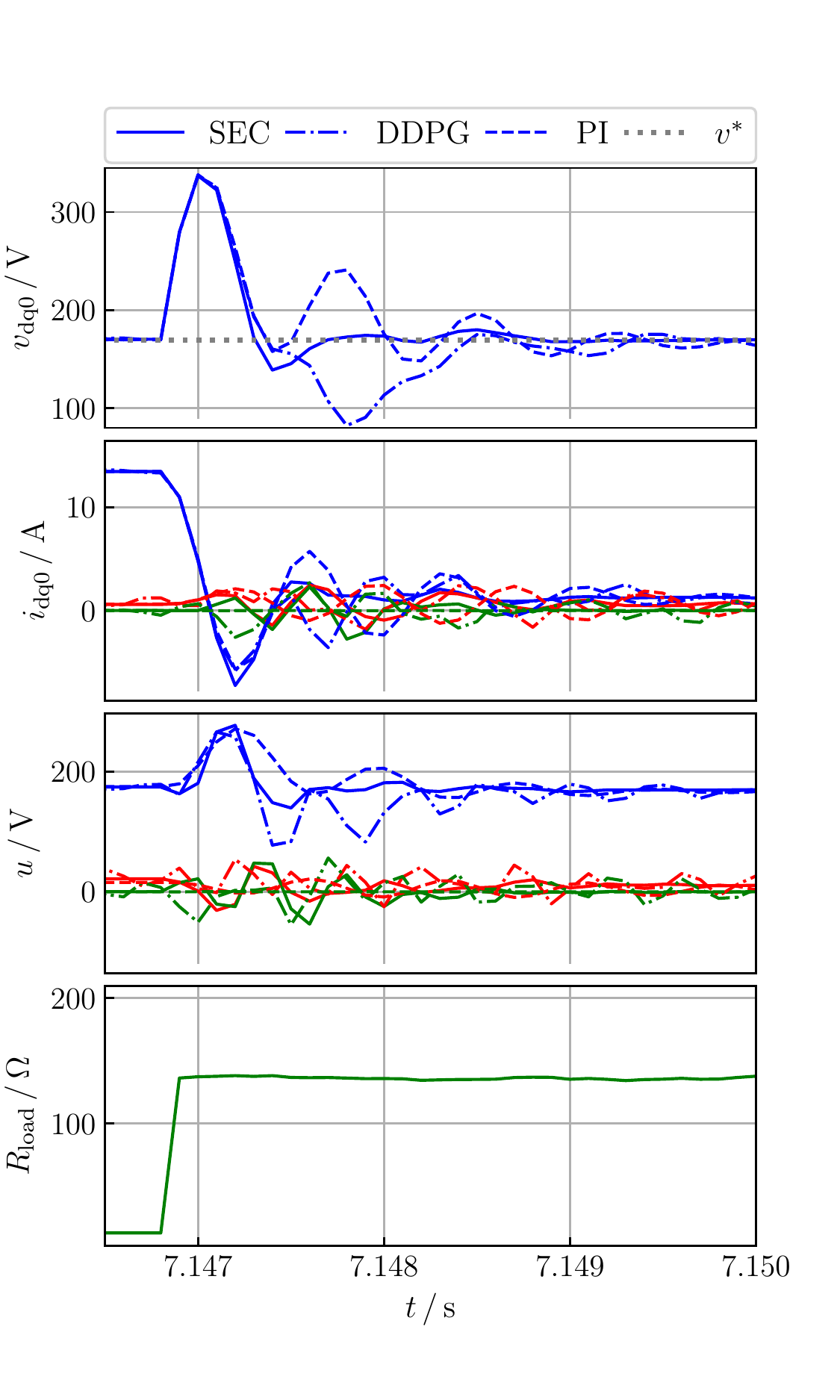}
\caption{Exemplary loadstep 1}
\label{fig:OMG_DDPGpv_PI_compare1}
\end{subfigure}
\begin{subfigure}{.3\textwidth}
\centering
\includegraphics[width=1\linewidth]{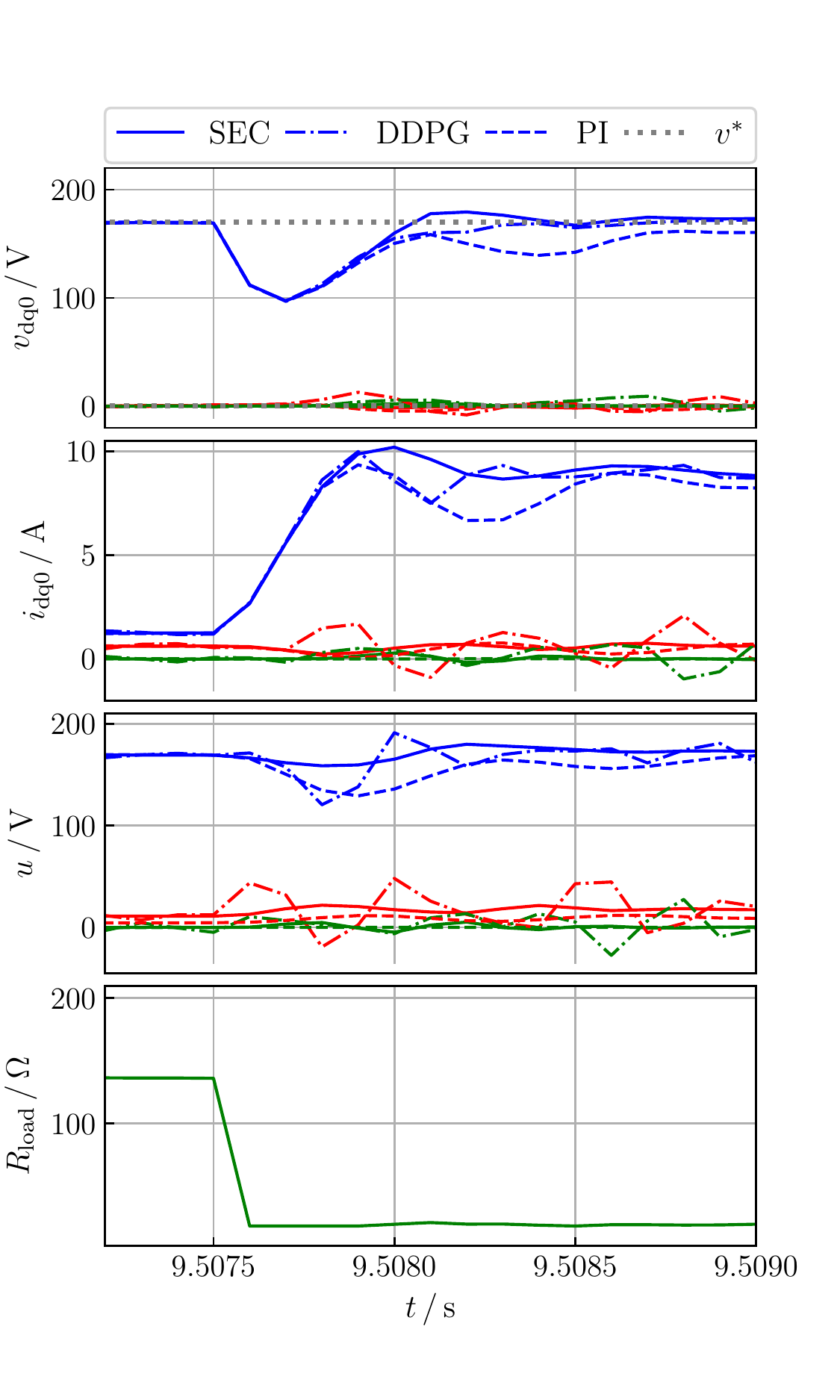}
\caption{Exemplary loadstep 2}
\label{fig:OMG_DDPGpv_PI_compare2}
\end{subfigure}
\begin{subfigure}{.3\textwidth}
\centering
\includegraphics[width=1\linewidth]{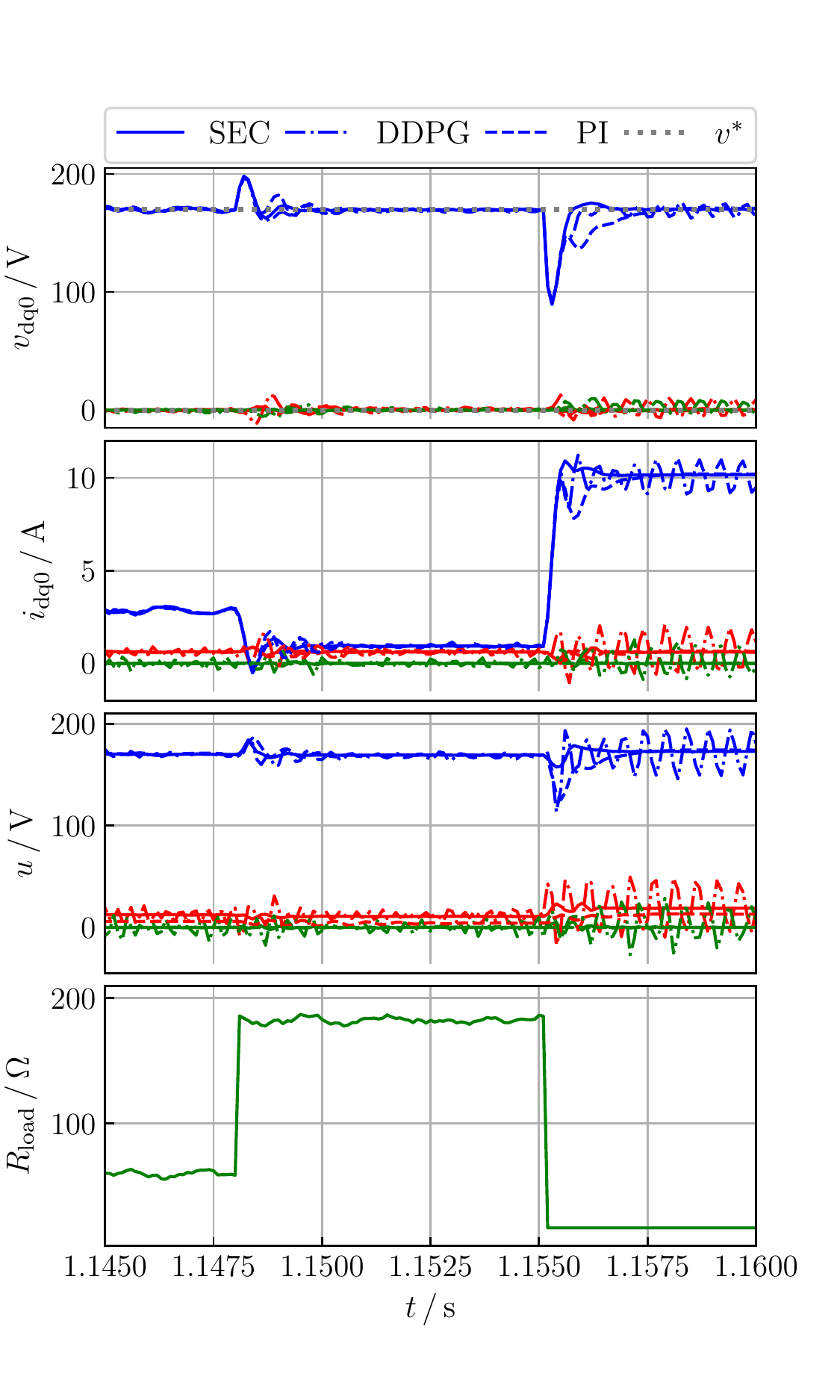}
\caption{Exemplary loadstep 3}
\label{fig:OMG_DDPGpv_PI_compare3}
\end{subfigure}
\caption{Qualitative comparison of the best RL control approach with / without SEC and the optimized PI controller for transients in the power grid application showing \textcolor{blue}{$v_\mathrm{d}$}, \textcolor{red}{$v_\mathrm{q}$} and \textcolor{mygreen}{$v_\mathrm{0}$}. The reference values are represented by the grey dotted lines.}
\label{fig:OMG_DDPGpv_PI_compare}
\end{figure*}
\subsection{Performance Analysis}
In the following the behavior of the best performing SEC-DDPG agent is compared to the best performing standard DDPG agent and to a standard PI control approach for both exemplary applications.
\subsubsection{Transient Behavior  -- Power Grid Application}
Firstly, the power grid application described in \capref{cap:OMG_setting} is investigated.
The PI controller, implemented as a cascade of voltage and current controllers, is parameterized using safe Bayesian optimization \cite{Weber2021} over 100 iterations.
More information about the parameterization  can be found in \cite{Weber2021a}.
%
%
%
%
%
In \figref{fig:OMG_DDPGpv_PI_compare}, three exemplary qualitative timeseries comparisons between the SEC-DDPG, standard DDPG and the optimized PI controller are presented during changes of the disturbance load current in the power grid scenario.
As described in \capref{cap:DDPH_HPs}, the controller performance concerning the given task is evaluated with use of the task-specific reward \eqref{eq:omg_reward_v} (setting $\gamma = 0$ for the evaluation). 
In both, the transients depicted in \figref{fig:OMG_DDPGpv_PI_compare} as well as in the validation episode specified by the load behavior in \figref{fig:OMG_R_load}, the SEC-DDPG agent surpasses the performance of the PI controller as well as the standard DDPG agent 
and achieves a better average reward 
as listed in \tabref{tab:OMG_results}. 
It can be seen that the SEC-DDPG agent is able to ensure stationary accuracy while also reacting faster during transients.
\subsubsection{Steady-State Behavior -- Power Grid Application}
To validate the disturbance rejection performance within the power system scenario, a $1\,\mathrm{s}$ test case with step-wise constant load values is generated using the DESSCA algorithm described in \capref{cap:GEM_setting} \cite{Schenke2021a}. 
The load value is changed every $500$ steps.
The only time-variable disturbance that occurs within the $500$ steps is the measurement noise provided by the environment (compare \cite{Weber2021a}).
To ensure that only the steady-state is considered for evaluation, the first $100$ of these $500$ steps are neglected to account for the transient phase.
The same metric as described above is applied to the remaining $400$ steps to evaluate the steady-state behavior of the different control approaches.
In \tabref{tab:OMG_results} the average performance over all $20$ constant load cases is presented.
It can be concluded that the steady-state behavior of the DDPG algorithm is improved by more than \unit{50}{\%} applying the IASA.
Furthermore, the SEC approach is performing slightly better than the PI controller as listed in \tabref{tab:OMG_results}.
\begin{figure}[htb]
\centering
\includegraphics[width=1\linewidth]{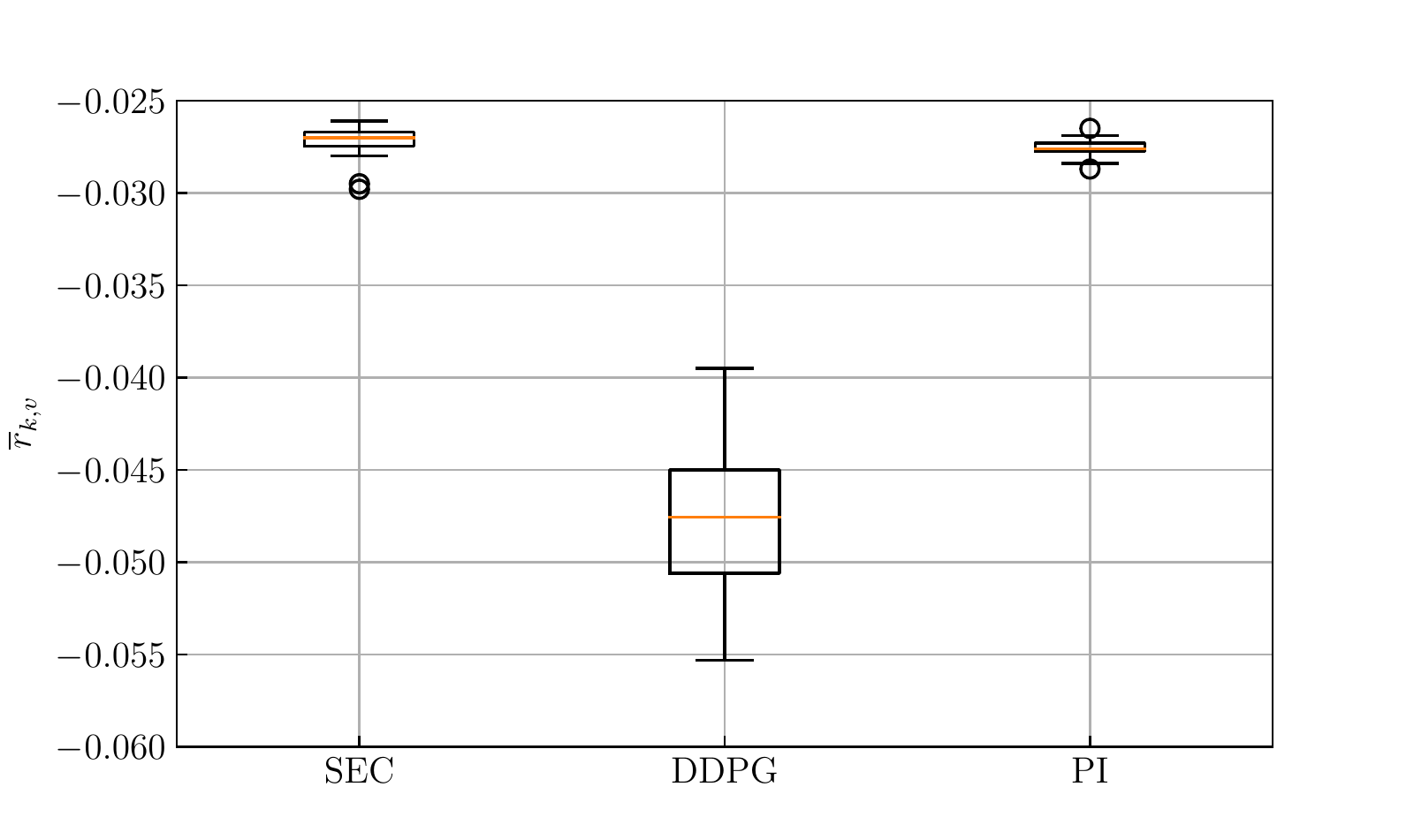}
\caption{Box plots of the control performance distribution evaluated by the mean reward during $20$ steady-state load cases in the power grid scenario. For the SEC and DDPG the best control agents based on the test case in Fig.~\ref{fig:OMG_R_load} are reported.}
\label{fig:OMG_Errorbar_SSE}
\end{figure}
%
%
%
%
%
\begin{table}[ht]
  \caption{Comparison of average reward using \eqref{eq:omg_reward_v} (with \mbox{$\gamma = 0$}) the best standard DDPG, best SEC-DDPG agent and PI controller in the power grid application.}
  \label{tab:OMG_results}
  \centering
  \begin{tabular}[t]{ l | c c c }
    \textbf{} & $\overline{r}_{\mathrm{k}, v, \mathrm{SEC}}$ & $\overline{r}_{\mathrm{k}, v, \mathrm{DDPG}}$ & $\overline{r}_{\mathrm{k}, v, \mathrm{PI}}$\\
    \hline
    \figref{fig:OMG_DDPGpv_PI_compare1} -- loadstep & $-0.1256$ & $-0.2073$ & $-0.1469$\\
    \figref{fig:OMG_DDPGpv_PI_compare2} -- loadstep & $-0.0916$ & $-0.1264$ & $-0.1157$ \\
    \figref{fig:OMG_DDPGpv_PI_compare3} -- loadstep & $-0.0437$ & $-0.0841$ & $-0.0562$ \\
    \figref{fig:OMG_R_load} -- $10 \, \second$ test case & $-0.0322$ & $-0.0572$ & $-0.0332$\\
    \figref{fig:OMG_Errorbar_SSE} -- steady state & $-0.0273$ & $-0.0528$ & $-0.0276$\\
    (mean out of 20 runs) & & &
  \end{tabular}
\end{table}
\begin{figure*}[htb]
\centering
\begin{subfigure}{.3\textwidth}
\centering
\includegraphics[width=1\linewidth]{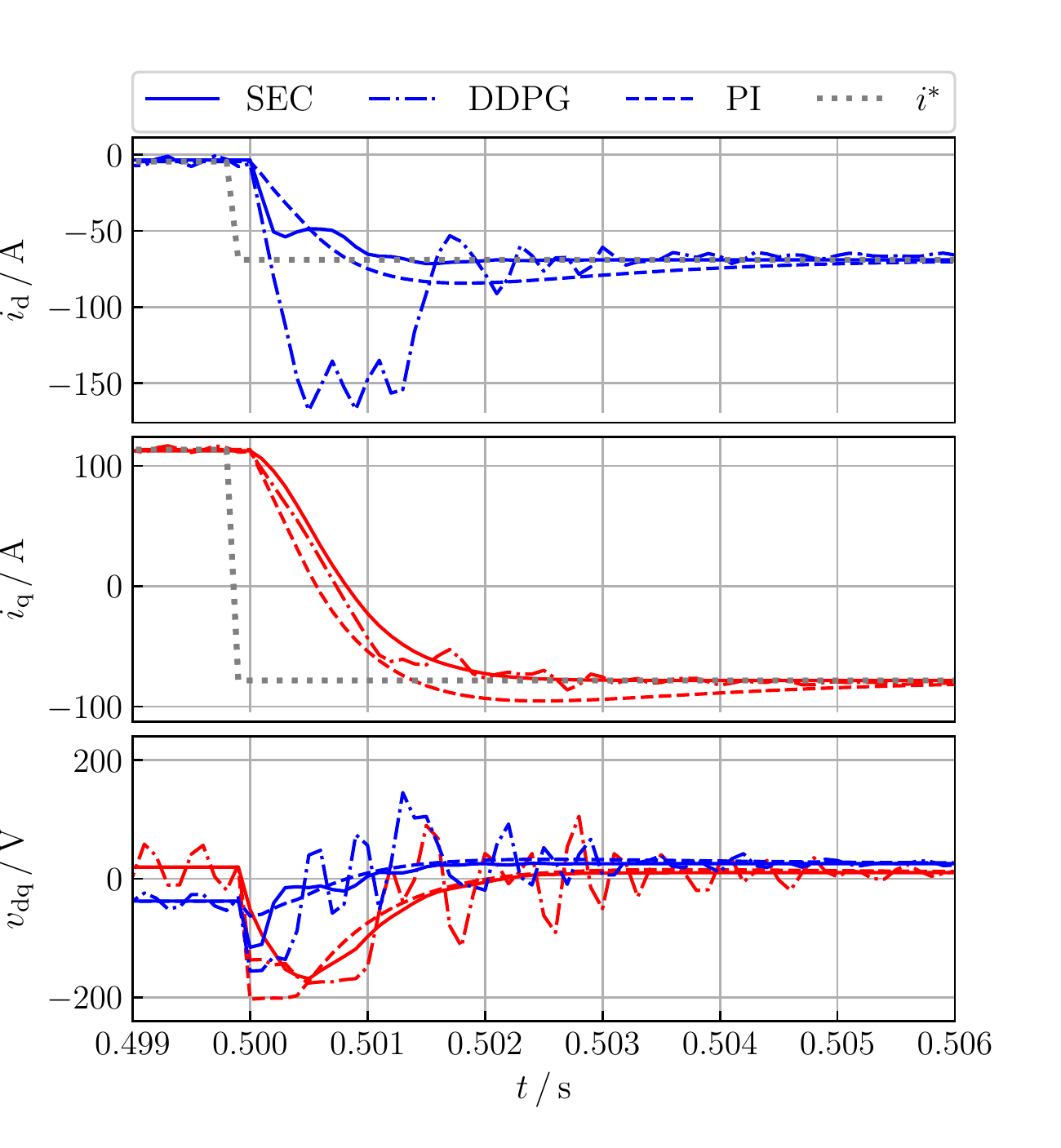}
\caption{Exemplary reference change 1}
\label{fig:GEM_DDPG_I_noI_idq1}
\end{subfigure}
\begin{subfigure}{.3\textwidth}
\centering
\includegraphics[width=1\linewidth]{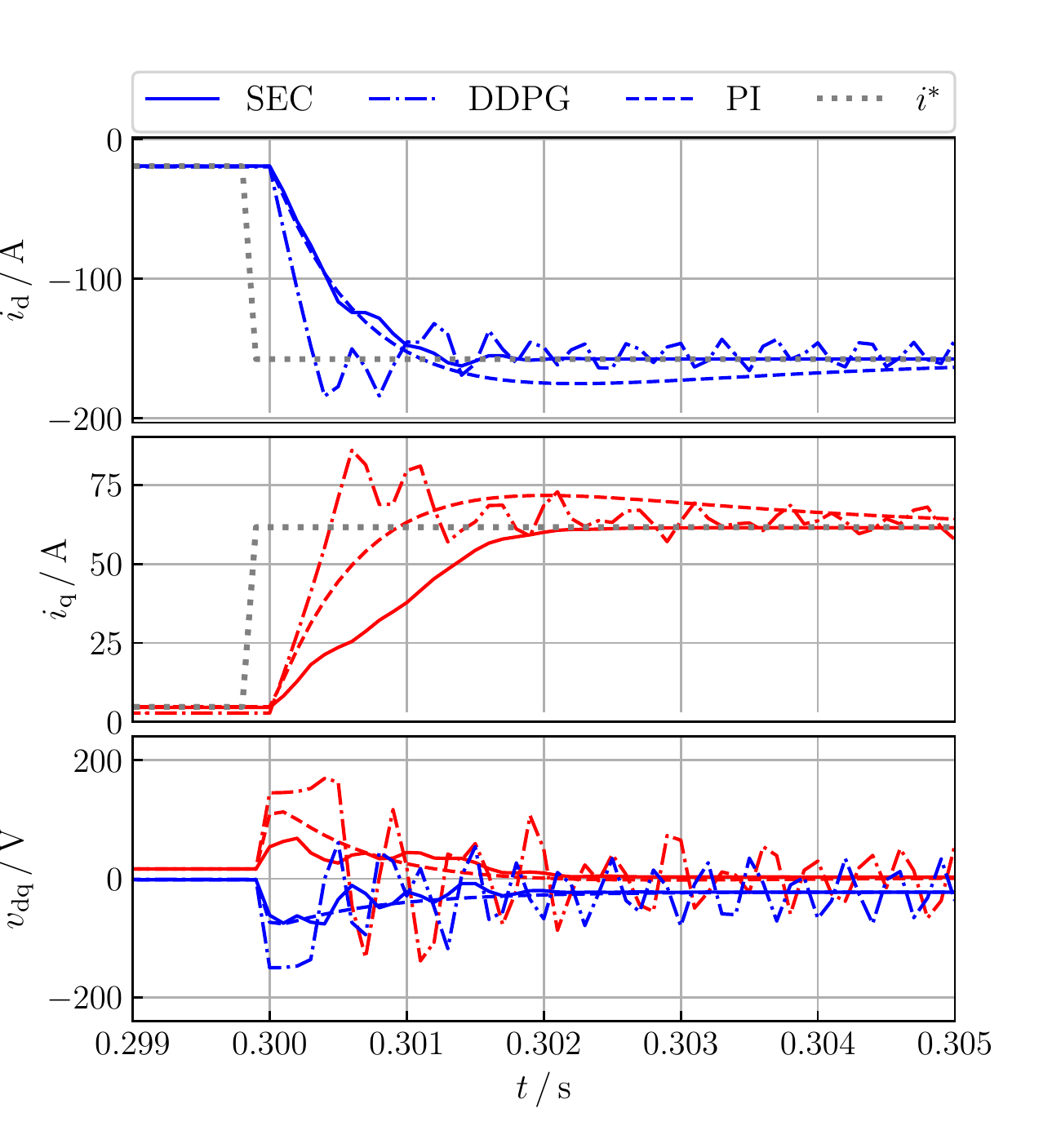}
\caption{Exemplary reference change 2}
\label{fig:GEM_DDPG_I_noI_idq2}
\end{subfigure}
\begin{subfigure}{.3\textwidth}
\centering
\includegraphics[width=1\linewidth]{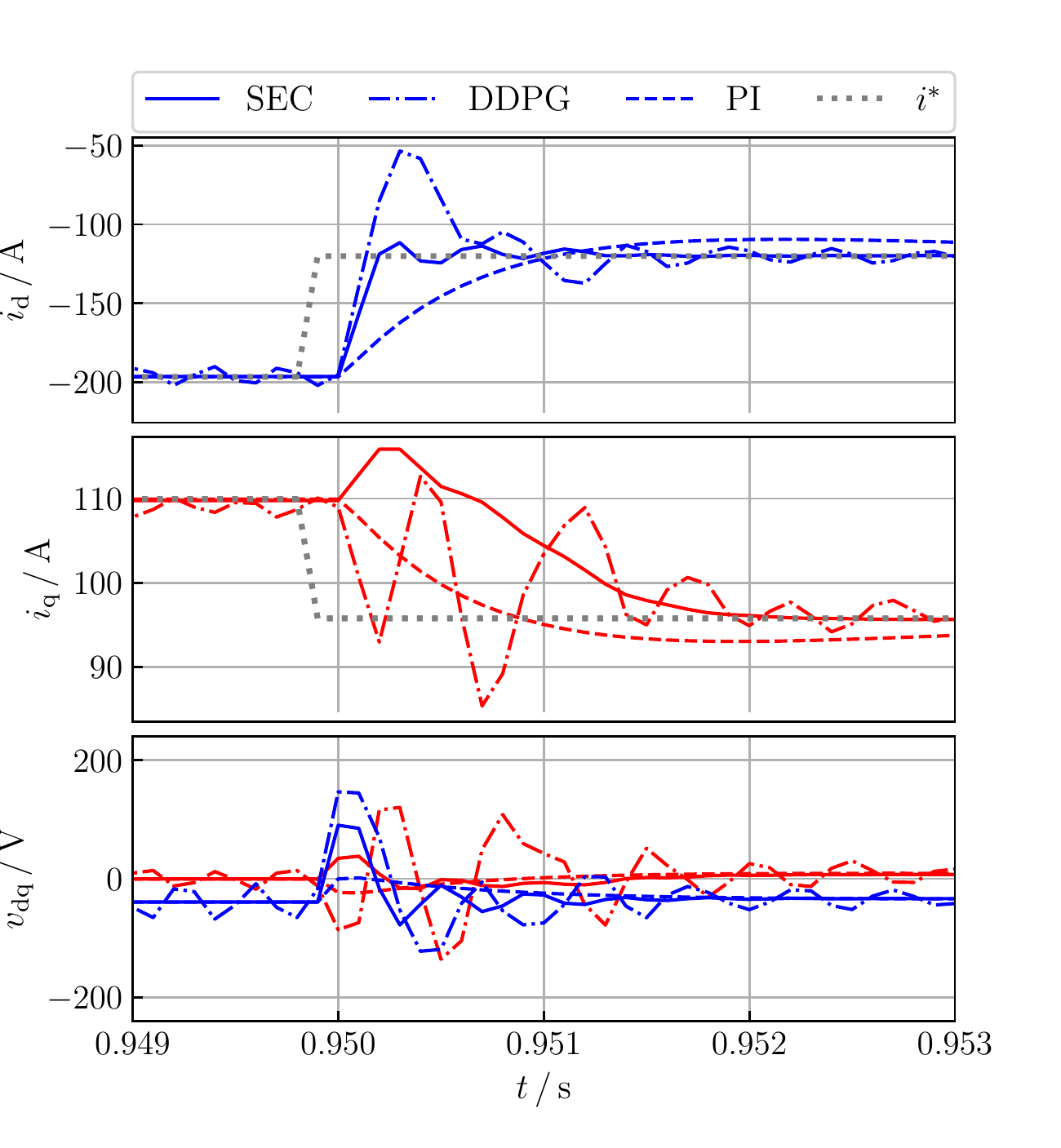}
\caption{Exemplary reference change 3}
\label{fig:GEM_DDPG_I_noI_idq3}
\end{subfigure}
\caption{Qualitative comparison of the DDPG controller with / without SEC in the electric drive application. The reference values are represented by the grey dotted lines.}
\label{fig:GEM_DDPG_I_noI_idq}
\end{figure*}

\begin{table}[htb]
  \caption{Comparison of average reward using \eqref{eq:gem_reward_i} (with \mbox{$\gamma = 0$}) for the best standard DDPG, best SEC-DDPG agent and PI controller in the electric drive application.}
  \label{tab:GEM_results}
  \centering
  \begin{tabular}[t]{ l | c c c}
    \textbf{}&$\overline{r}_{\mathrm{k}, i, \mathrm{SEC}}$&$\overline{r}_{\mathrm{k}, i, \mathrm{DDPG}}$ &$\overline{r}_{\mathrm{k}, i, \mathrm{PI}}$\\\hline
    \figref{fig:GEM_DDPG_I_noI_idq1} -- changing reference & $-0.1948$ & $-0.3146$ &$-0.2179$\\
    \figref{fig:GEM_DDPG_I_noI_idq2} -- changing reference & $-0.1891$ & $-0.2496$ &$-0.2071$\\
    \figref{fig:GEM_DDPG_I_noI_idq3} -- changing reference & $-0.1437$ & $-0.2254$ & $-0.1451$\\
    \capref{cap:GEM_setting} -- $1 \, \second$ test case & $-0.0745$ & $-0.1443$ & $-0.0371$\\
    \figref{fig:GEM_Errorbar_SSE} -- steady state& $-0.0473$ & $-0.1236$ & $-0.0002$\\
    (mean out of 20 runs) & & &
  \end{tabular}
\end{table}
%
%
%
%
%
\begin{figure}[htb]
\centering
\includegraphics[width=1\linewidth]{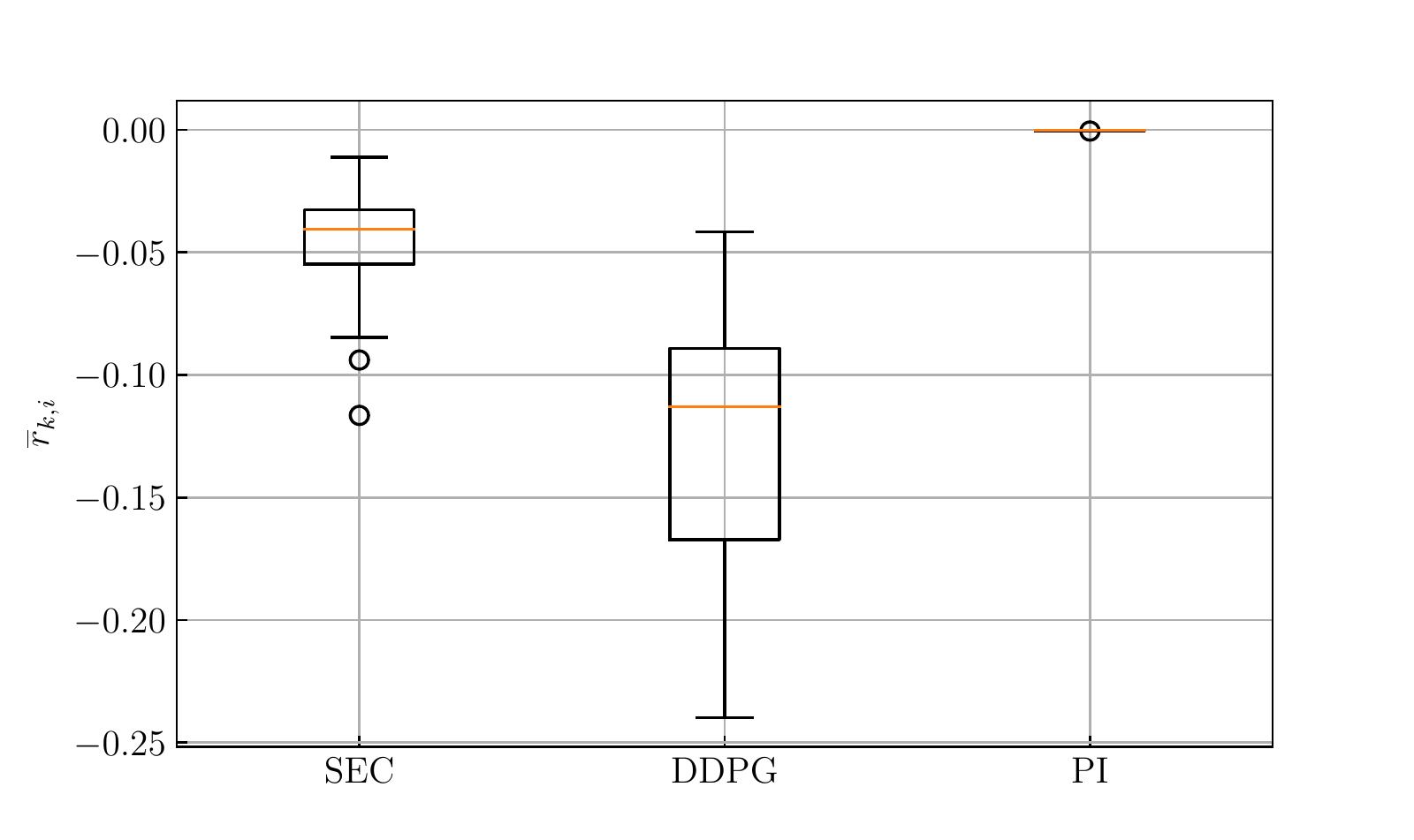}
\caption{Box plots of the control performance distribution evaluated by the mean reward during during $20$ steady-state load cases in  the electric drive scenario. For the SEC and DDPG the best control agents based on the test case described in \capref{cap:drive_training} are reported.}
\label{fig:GEM_Errorbar_SSE}
\end{figure}
\subsubsection{Transient Behavior  -- Electric Motor Application}
Further, the electric motor application described in \capref{cap:GEM_setting} is utilized to qualitatively compare the best standard DDPG, the best SEC-DDPG agent and a PI controller.
The controller parameters of the PI controller are determined utilizing the symmetrical optimum.
Again, more information about the parameterization  can be found in \cite{Weber2021a}.
The average reward for this task (as defined in \eqref{eq:gem_reward_i} and setting $\gamma = 0$) is used as performance metric.
In \figref{fig:GEM_DDPG_I_noI_idq}, the performance of the current control task is depicted during different reference changes. 
It is visible that the SEC-DDPG agent is able to ensure stationary accuracy while reacting as fast as the standard DDPG agent during transient behavior, whereas the standard DDPG agent is not able to fully compensate the steady-state error. 
Compared to the PI controller it can be observed that the SEC-DDPG agent reaches the steady state faster and with less overshoot.
As listed in \tabref{tab:GEM_results}, the SEC-DDGP agent performs best in the presented reference step episodes.
Within the $1 \, \second$ test case described in \capref{cap:drive_training}, the PI controller performs best.
It can be expected that the applied hyperparameters, which have been fitted for the power grid application, would still allow a noticeable improvement of the SEC-DDPG's performance if optimized for the electric motor application.
Nonetheless, the applicability of the suggested method can be confirmed for this environment.
\subsubsection{Steady-State Behavior  -- Electric Motor Application}
A similar evaluation is performed to validate the stationary precision within the electric motor application.
Again, the $1 \, \mathrm{s}$ test case scenario with a stepwise constant reference is considered, wherein one reference value is kept for 500 timesteps.
In order to account for slower dynamics of the system, $200$ timesteps are neglected after each reference step to neglect the transients.
In \tabref{tab:GEM_results} the average performance over the $20$ references is listed for the three different control approaches. 
As visible in \figref{fig:GEM_Errorbar_SSE}, the PI controller performs best, which is not surprising since no disturbance was present.
The DDPG's performance, however, was surpassed by more than \unit{38}{\%} using the SEC extension.

%
%


%% file: Conclusion.tex
\section{Conclusion and Outlook}
\label{cap:conclusion+}
In the applications under investigation it has been demonstrated that standard RL algorithms have difficulties reducing the steady-state error in reference and disturbance rejection tracking control problems. 
The introduced SEC method for RL adds an integral style feedback to the actor output, which reduces the steady-state error significantly.
Within a control problem with set-point reference and significant disturbance the augmented RL controller performed better in comparison to an optimized PI controller.
The SEC-RL controller was able to sufficiently handle the steady-state error while featuring better performance during transients. In a control scenario with changing reference values, the SEC approach was also able to drastically improve the control performance in comparison to a default RL algorithm.

Future investigations on this topic should include the experimental validation on real-world test benches, where further parasitic effects are to be handled.
Moreover, the training routine should be further investigated to find modifications reducing the learning variance in the SEC approach.
In addition, different approaches to the reward design and their impact on the performance during transients and steady-state operation can be investigated.

%% file: acknowledgment.tex
\section{Acknowledgment}
The authors gratefully acknowledge the funding of this project by computing time provided by the Paderborn Center for Parallel Computing $(\text{PC}^2)$.